\documentclass[prb, twocolumn, superscriptaddress, nofootinbib, floatfix, aps, 10pt]{revtex4-2}
\usepackage{graphicx}
\usepackage{float}
\usepackage{soul}
\usepackage[normalem]{ulem}
\usepackage{amsmath,amssymb,amstext,dsfont,tikz,graphicx,physics,mathtools,bm,
simpler-wick,stmaryrd}
\usepackage{overpic}
\usepackage[caption=false]{subfig}
\usepackage[colorlinks=true, allcolors=purple]{hyperref}

\makeatletter 
    
\renewcommand\onecolumngrid{
\do@columngrid{one}{\@ne}%
\def\set@footnotewidth{\onecolumngrid}
\def\footnoterule{\kern-6pt\hrule width 1.5in\kern6pt}%
}

\renewcommand\twocolumngrid{
        \def\footnoterule{
        \dimen@\skip\footins\divide\dimen@\thr@@
        \kern-\dimen@\hrule width.5in\kern\dimen@}
        \do@columngrid{mlt}{\tw@}
}%

\makeatother

\newcommand{\hclass}[1]{\llbracket #1 \rrbracket}

\DeclareMathOperator{\el}{el}
\DeclareMathOperator{\AF}{AF}
\DeclareMathOperator{\FS}{FS}
\DeclareMathOperator{\stag}{stag}
\DeclareMathOperator{\RBIM}{RBIM}

\DeclareMathOperator{\MW}{MW}
\DeclareMathOperator{\opt}{opt}

\begin{document}

\title{Computational Phase Transitions in Two-Dimensional Antiferromagnetic Melting}

\author{Zack Weinstein}
\email{zackmweinstein@berkeley.edu}
\affiliation{Department of Physics, University of California, Berkeley, CA 94720, USA}

\author{Jalal Abu Ahmad}
\affiliation{Physics Department, Technion, 32000 Haifa, Israel}

\author{Daniel Podolsky}
\affiliation{Physics Department, Technion, 32000 Haifa, Israel}

\author{Ehud Altman}
\affiliation{Department of Physics, University of California, Berkeley, CA 94720, USA}
\affiliation{Materials Sciences Division, Lawrence Berkeley National Laboratory, Berkeley, CA 94720, USA}

\date{\today}

\begin{abstract}
A computational phase transition in a classical or quantum system is a non-analytic change in behavior of an order parameter which can only be observed with the assistance of a nontrivial classical computation. Such phase transitions, and the computational observables which detect them, play a crucial role in the optimal decoding of quantum error-correcting codes and in the scalable detection of measurement-induced phenomena. In this work we show that computational phase transitions and observables can also provide important physical insight on the phase diagram of a classical statistical physics system, specifically in the context of the dislocation-mediated melting of a two-dimensional antiferromagnetic (AF) crystal. In the solid phase, elementary dislocations disrupt the bipartiteness of the underlying square lattice, and as a result, pairs of dislocations are linearly confined by string-like AF domain walls. It has previously been argued that a novel AF tetratic phase can arise when double dislocations proliferate while elementary dislocations remain bound. However, since elementary dislocations carry AF Ising gauge flux, no local order parameter can distinguish between AF and paramagnetic (PM) tetratic regimes, and consequently there is no thermodynamic phase transition separating the two regimes. Nonetheless, we demonstrate that it is possible to \textit{algorithmically} construct a staggered magnetization which distinguishes the AF and PM tetratic regimes by ``pairing'' dislocations, which requires an increasingly nontrivial classical computation as elementary dislocation pairs increase in density and unbind. We discuss both algorithm-dependent and ``intrinsic'' algorithm-independent computational phase transitions in this setting, the latter of which includes a transition in one's ability to consistently sort atoms into two sublattices to construct a well-defined staggered magnetization.

\end{abstract}

\maketitle

\section{Introduction}
The modern theory of classical and quantum critical phenomena has been pushed far beyond Landau's original symmetry-breaking paradigm \cite{landau1937theory,goldenfeldLecturesPhaseTransitions1992}. While traditional symmetry-broken phases are distinguished from each other by the behavior of \textit{local} order parameters, much of the significant progress in describing beyond-Landau phases of matter, and the phase transitions between them, has arisen from the study of increasingly sophisticated and often \textit{nonlocal} observables. For example, zero-dimensional point-like order parameters are generalized to the one-dimensional Wilson loops of lattice gauge theories \cite{wegner1971duality,wilson1974confinement,kogut1979introduction}, and to open-string order parameters in symmetry-protected topological phases \cite{denNijs1989Preroughening,DallaTorre2006Hidden,PerezGarcia2008String,Berg2008Rise,Pollmann2012Symmetry,verresen2024higgs}. An even larger departure from the traditional notion of an order parameter has been found in \textit{information-theoretic} observables; in the quantum setting, quantum entanglement measures have proven invaluable in characterizing conformal field theories \cite{holzhey1994geometric,calabrese2009entanglement,pollmann2009theory}, topological order \cite{kitaev2006topological,levin2006detecting,chen2010local,jiang2012identifying,lu2020detecting}, and far-from-equilibrium dynamical phases \cite{bardarson2012unbounded,nanduri2014entanglement,dumitrescu2017scaling,lukin2019probing,nahum2017quantum,li2018quantum,li2019measurement,skinner2019measurement,bao2020theory,jian2020measurement}, while in the classical realm, the R\'{e}nyi mutual information has been demonstrated to detect phase transitions without prerequisite knowledge of a traditional order parameter \cite{iaconisDetectingClassicalPhase2013}. 

The present work is concerned with an even broader class of observables called \textit{computational} observables. These are observables of a classical or quantum system which generally require a nontrivial classical computation to measure or calculate. Computational observables may be defined either by a specific algorithm, or by an ``intrinsic'' property of the system which cannot efficiently be probed by conventional observables. In classical statistical physics, one example of such a computational observable is given by a recent worldsheet patching algorithm for $\mathbb{Z}_2$ lattice gauge theories with matter \cite{somoza2021self,serna2024worldsheet}, which is used to construct a dual order parameter which does not exist at the microscopic level, but is expected to emerge at larger length scales. In a similar vein, Ref.~\cite{congEnhancingDetectionTopological2024} proposed and implemented a practical method for improving Wilson loop measurements in experimental realizations of topologically ordered states, using a renormalization group inspired decoding protocol.  In a very different setting, computational observables such as ``quantum-classical cross-correlations" have been recognized as perhaps the only scalable probes of measurement-induced phenomena in many-body systems \cite{Gullans2020Scalable,Li2023Decodable,noelMeasurementinducedQuantumPhases2022,Barratt2022Transitions,Garratt2023Measurements,lee2022decoding,Li2023Cross,dehghaniNeuralnetworkDecodersMeasurement2023,hokeMeasurementinducedEntanglementTeleportation2023,garratt2023probing,mcginley2023postselectionfree}. A common feature of many of these computational observables is a close analogy to protocols developed originally for quantum error correction \cite{dennisTopologicalQuantumMemory2002,bombinIntroductionTopologicalQuantum2013,nielsenQuantumComputationQuantum2010}: although the desired physical behavior cannot be immediately seen in simple expectation values, it can be ``decoded'' by first performing a series of corrections on the underlying state. 

In this work, we show that computational observables can be important for making sense of observations in classical statistical physics and point to a sharp distinction between computational phase transitions and usual thermodynamic phase transitions. Specifically, we identify non-thermodynamic computational phase transitions which arise naturally in models of classical two-dimensional melting in the presence of Ising antiferromagnetism. These computational transitions, and the computational order parameters which detect them, are necessary to precisely explain the qualitative physics observed in previous studies of antiferromagnetic melting \cite{abutbulTopologicalOrderAntiferromagnetic2022}, which is not properly captured by any ordinary thermodynamic phase transitions or local observables.

The system we consider consists of a collection of atoms, each endowed with a positional coordinate $\vb{r}_i$ and an Ising spin $\sigma_i$, which interact both elastically and antiferromagnetically. In the absence of antiferromagnetism, classical melting in two spatial dimensions is famously described by Kosterlitz-Thouless-Halperin-Nelson-Young (KTHNY) theory \cite{kosterlitzLongRangeOrder1972,kosterlitzOrderingMetastabilityPhase1973a,halperinTheoryTwoDimensionalMelting1978a,nelsonDislocationmediatedMeltingTwo1979b,youngMeltingVectorCoulomb1979a,chaikinPrinciplesCondensedMatter2013}, in which finite-temperature translational quasi-long-range order (QLRO) is destroyed by the proliferation of pointlike topological defects called \textit{dislocations}. When the atoms are also given Ising antiferromagnetic (AF) interactions, previous works \cite{cardyTransitionsInternalDegree1983,abutbulTopologicalOrderAntiferromagnetic2022} have pointed out a remarkable interplay between dislocations and AF order on the resulting deformed lattice. Namely, elementary dislocations frustrate the underlying AF order by disrupting the lattice's bipartiteness, and as a result, each elementary dislocation is necessarily bound to the endpoint of a string-like Ising domain wall [see Fig.~\ref{fig:dislocations}(b)]. In the presence of a strong AF interaction strength $J$, this phenomenon leads to a \textit{linear} confinement of elementary dislocations, a strong enhancement over the logarithmic interactions between dislocations arising from the solid's elastic rigidity. In contrast, \textit{pairs of} double-dislocations experience no such linear confinement. When the AF interactions are sufficiently strong, the first topological defects to proliferate as the temperature is raised are \textit{double} dislocations. 

If elementary dislocations are completely absent when double dislocations proliferate, an unusual \textit{antiferromagnetic tetratic} phase is realized \cite{abutbulTopologicalOrderAntiferromagnetic2022}. While double dislocations erode the lattice's quasi-long-range translational order, the bipartiteness of the lattice remains undisturbed; i.e., so long as the typical distance between double dislocations remains appreciably larger than the inter-atomic spacing, the atoms can be consistently sorted into two interpenetrating sublattices\footnote{Throughout this work, we often use the words ``lattice" and ``sublattice" even in the absence of positional (quasi)-long-range order. This terminology is appropriate so long as the positional correlation length is much larger than the inter-atomic spacing, so that atoms are \textit{locally} organized into a square lattice.}, allowing in principle for unfrustrated long-range AF order despite the short-range positional order. In the absence of a fixed lattice structure, this long-range AF order can be diagnosed by a staggered magnetization (i.e., the difference in magnetizations of the two opposing sublattices), or by a string correlation function which counts whether two far-separated atoms sit on the same or opposing sublattices. Intuitively, we might naively expect this long-range AF order to be robust to a dilute density of tightly-bound elementary dislocation pairs. Indeed, large-scale Monte Carlo simulations suggest that strong AF correlations and a nearly bipartite lattice can survive within the tetratic phase \cite{abutbulTopologicalOrderAntiferromagnetic2022,jalal_thesis}.

These features of the AF tetratic are in sharp contrast to those of the conventional paramagnetic (PM) tetratic, which can be obtained from the AF solid by first destroying the AF order to reach a PM solid, and then proliferating single dislocations. In the PM tetratic, a finite density of free elementary dislocations destroys any notion of an approximate bipartite lattice. As a result, not only are the obvious signatures of long-range AF order absent, but it is not even clear how such observables should be defined or computed! This situation is starkly different from that of ordinary symmetry-breaking phase transitions, where an easily computable order parameter is well-defined in both phases and its \textit{value} is used to distinguish between the different phases. It is therefore interesting to ask how AF order should be characterized as one interpolates between the AF and PM tetratics, and specifically whether and how one can define an order parameter which can sharply distinguish between these two regimes.

We will show in this work that the AF and PM tetratics cannot be distinguished by any ordinary \textit{local} order parameter; in fact, the two regimes are adiabatically connected to each other within the same thermodynamic phase. Nevertheless, one can construct a \textit{nonlocal} order parameter for the AF tetratic regime by algorithmically pairing dislocations together, thereby bipartitioning the atoms into two sublattices. This pairing procedure can be thought of as a form of ``error correction'' performed on the lattice, and is closely analogous to quantum error correction protocols in surface codes \cite{dennisTopologicalQuantumMemory2002,bombinIntroductionTopologicalQuantum2013}. The staggered magnetization and string correlation functions resulting from this bipartitioning are examples of \textit{computational} observables. We will demonstrate that these observables are capable of detecting non-thermodynamic computational phase transitions which separate the AF and PM tetratic regimes.

First, let us explain why the AF and PM tetratic regimes are not separated by a sharp thermodynamic phase transition, but instead by a smooth crossover. Since elementary dislocations are bound to the endpoints of string-like Ising domain walls, they can be thought of as carrying Ising gauge flux. Within the tetratic phase, the spin degrees of freedom are thus best understood not as an Ising model, but as a \textit{gauged} Ising model. The AF and PM tetratic regimes correspond respectively to the Higgs and confined regimes of this Ising gauge theory, which are well-known to be adiabatically connected to each other \cite{fradkinPhaseDiagramsLattice1979}. This observation, originally made by Ref.~\cite{cardyTransitionsInternalDegree1983}, explains the difficulty in constructing an order parameter which distinguishes between the AF and PM tetratic regimes: as the AF interaction strength is reduced, the presence of nontrivial gauge flux throughout the system acts as an obstruction to globally defining an Ising order parameter, allowing for a crossover between the two regimes without the need for a symmetry-breaking phase transition.

The gauge theory perspective suggests that, at very large scales, true long-range AF order is eventually destroyed in the tetratic phase by elementary dislocations. However, this theoretical observation seemingly contradicts the results of recent large-scale Monte Carlo simulations \cite{jalal_thesis}, which observe near-perfect AF order within the tetratic phase of systems with as many as 90,000 particles. In these simulations, since elementary dislocations almost always arise in tightly-bound pairs for large $J$, a bipartite lattice and well-defined AF observables are easily recovered by systematically ``ignoring'' these bound pairs. The resulting staggered magnetization, defined simply by subtracting the magnetizations of the two resulting sublattices, is observed to be nearly maximal and non-decaying with increasing system size. More generally, a well-defined bipartitioning of the atoms can always be recovered by \textit{pairing} dislocations, which is performed by drawing paths through the dual lattice which connect the atoms pairwise. Neglecting potential global issues to be discussed in detail below, each such pairing defines a bipartitioning of the atoms by allowing nearest-neighbor atoms which are bisected by these paths to belong to the same sublattice. AF domains are defined simply as the regions bounded by both the physical domain walls and the pairing paths, which together form closed loops.

The process of pairing dislocations is relatively local for large $J$, but it becomes increasingly nonlocal as $J$ is reduced. Although it quickly becomes difficult to decide on a pairing of dislocations by hand, a classical algorithm can be introduced to systematically pair dislocations well-beyond the point at which AF order becomes visually unrecognizable. For example, one natural prescription for pairing dislocations is via a minimal-weight matching algorithm \cite{edmondsPathsTreesFlowers1965,barahonaMorphologyGroundStates1982}, which minimizes the number of nearest-neighbor atom pairs which are bipartitioned into the same sublattice. Since a staggered magnetization defined this way is large in magnitude for large $J$ and uniformly zero for small $J$, we expect the AF and PM tetratic regimes to be separated by a non-thermodynamic \textit{computational phase transition}, characterized by a non-analytic change in behavior of the classical algorithm used to define these observables. In particular, we expect the Ising domains defined by this dislocation-pairing algorithm to largely consist of a single infinite cluster in the AF tetratic regime, while in the PM tetratic regime no infinite cluster is created by pairing dislocations.

The computational observables defined via dislocation-pairing are somewhat algorithm-dependent, and in principle the location of the computational transition depends on the choice of algorithm used. Ideally, we would like to draw an \textit{intrinsic}, \textit{algorithm-independent} distinction between the AF and PM tetratic regimes. One way this can be accomplished is by considering the model in the presence of periodic boundary conditions, where our dislocation-pairing protocol can exhibit a particularly interesting mode of failure. Specifically, when the system is placed in a space with nontrivial topology, the possible domain wall configurations for a given set of dislocation positions can fall into one of several inequivalent \textit{homology classes}. Two different domain wall configurations belong to the same homology class, and are called homologous, if one can be obtained from the other by flipping domains of Ising spins. If a given disloction pairing consistently bipartitions the atoms in one domain wall configuration, it will also consistently bipartition any homologous domain wall configuration. However, it will \textit{fail} to establish a consistent bipartitioning in any non-homologous domain wall configuration.

In the limit of large $J$, for each possible configuration of dislocations, only one of the possible homology classes will be statistically observable, while every other homology class arises with probability zero in the thermodynamic limit. Therefore, if this statistically guaranteed homology class can be determined for each possible configuration of dislocations, then it is always possible to establish a consistent bipartitioning of the atoms via dislocation-pairing. On the other hand, for small $J$, all homology classes will occur with comparable probabilities, and \textit{any} dislocation-pairing algorithm will simply fail to consistently define AF observables. As we shall show, these two regimes are separated by a computational phase transition. The AF and PM tetratic phases can therefore be sharply distinguished in the presence of topologically nontrivial boundary conditions by their ``bipartite-ability'', i.e., whether the atoms can be consistently bipartitioned so as to establish AF order parameters and correlation functions.

The remainder of this paper is organized as follows. In Sec.~\ref{sec:AFmelting} we define the primary model and outline its basic phenomenology, including its symmetries, order parameters, energetic excitations, and possible phases. In Sec.~\ref{sec:noPT}, we demonstrate the absence of a thermodynamic phase transition between the AF and PM tetratic regimes using a connection to Ising lattice gauge theory. The main results of our work are contained in Sec.~\ref{sec:computational}, where we show how computational observables can be algorithmically constructed to distinguish between the AF and PM tetratic regimes. In Sec.~\ref{subsec:comp_overview} we provide a broad overview of these computational observables, their expected behavior in important limiting regimes, and the types of non-thermodynamic computational phase transitions which can be detected. In Sec.~\ref{subsec:fs_mwpm}, \ref{subsec:fs_bipartiteness}, and \ref{subsec:optimal}, we demonstrate these ideas analytically and numerically in the simple and tractable setting of an Ising lattice gauge theory, which is expected to describe the behavior of the Ising spins within the tetratic phase to an excellent approximation. In Sec.~\ref{sec:hardspheres} we numerically demonstrate that dislocation-pairing algorithms can indeed construct well-defined AF observables in a microscopically realistic model of AF melting. Finally we discuss our results and some remaining questions in Sec.~\ref{sec:discussion}.

\section{Antiferromagnetic Melting}
\label{sec:AFmelting}

We consider a classical model of antiferromagnetic (AF) melting in two spatial dimensions. Specifically, our system consists of $N$ atoms in a two-dimensional box of linear size $L$, with both positional degrees of freedom $\vb{r}_i = (x_i, y_i)$ and Ising spins $\sigma_i = \pm 1$ ($i = 1, \ldots , N$). We leave the boundary conditions unspecified for the moment, although we are primarily interested in the thermodynamic limit $N, L \to \infty$ with the density of atoms $N / L^2$ fixed. Somewhat schematically, we can imagine that the atoms interact microscopically via a Hamiltonian of the form
\begin{equation}
\label{eq:ham_schematic}
	H[\vb{r},\sigma] = V[\vb{r}] + \frac{J}{2} \sum_{i \neq j}^N e^{1-\abs{\vb{r}_i - \vb{r}_j}/a} \sigma_i \sigma_j ,
\end{equation}
where $V[\vb{r}] \equiv V(\vb{r}_1, \ldots , \vb{r}_N)$ is an inter-atomic elastic potential energy which depends only on the magnitudes $\abs{\vb{r}_i - \vb{r}_j}$ of the relative positions of the atoms, while the second term provides a short-range Ising AF interaction between the atoms. Throughout this work we shall absorb the temperature and Boltzmann constant into the definition of $H$, so that thermodynamic expectation values are computed by sampling both atomic positions and spins according to the Boltzmann weight $e^{- H} / Z$:
\begin{equation}
	\expval{\cdots} = \frac{1}{Z} \tr \qty[ (\cdots) e^{- H} ], \quad Z = \tr e^{- H} ,
\end{equation}
where $\tr \equiv \sum_{\sigma} \int_{\vb{r}}$ denotes a sum over all spin configurations $\sigma = \qty{\sigma_i}$ and an integral over atomic positions $\vb{r} = \qty{\vb{r}_i}$.

The ground state of $H$ is simply the square-lattice N\'{e}el state, in which the atoms crystallize into a solid with lattice positions $\vb{r}_i = \vb{R}_{n_i, m_i} \equiv a(n_i \hat{\vb{x}} + m_i \hat{\vb{y}})$ and their spins $\sigma_i = (-1)^{n_i + m_i}$ alternate sign in a checkerboard pattern ($n_i, m_i \in \mathbb{Z}$). This state spontaneously breaks both the continuous translational symmetry $\vb{r}_i \mapsto \vb{r}_i + \vb{c}$ and the discrete Ising symmetry $\sigma_i \mapsto -\sigma_i$. A set of order parameters for the former symmetry-breaking are the Fourier modes of the atomic density at reciprocal lattice wavevectors $\vb{G}$:
\begin{equation}
	\rho_{\vb{G}} = \frac{1}{N} \sum_{i = 1}^N e^{-i \vb{G} \cdot \vb{r}_i}, \ \  \vb{G} = \frac{2\pi}{a} \qty( p \hat{\vb{x}} + q \hat{\vb{y}} ) , \ \ p, q \in \mathbb{Z}.
\end{equation}
An order parameter for the latter symmetry is the staggered magnetization, i.e., the $\vb{Q} = ( \frac{\pi}{a}, \frac{\pi}{a} )$ Fourier component of the magnetization density:
\begin{equation}
\label{eq:stag_mag}
	M_{\vb{Q}} = \frac{1}{N} \sum_{i = 1}^N \sigma_i e^{-i \vb{Q} \cdot \vb{r}_i} = \frac{1}{N} \sum_{i = 1}^N \sigma_i e^{-i \pi (x_i + y_i) / a} .
\end{equation}
In Sec.~\ref{sec:computational}, we will discuss another form of the staggered magnetization which does not explicitly involve the atomic positions, but generally must be defined algorithmically.

At low nonzero temperatures, the thermodynamics of $H$ can largely be described by three classes of energetic excitations:

\textit{(1) Phonons}: at nonzero temperatures, atoms can fluctuate about their equilibrium positions. Writing $\vb{r}_i = \vb{R}_{n_i, m_i} + \vb{u}_i$, and assuming that the displacements $\vb{u}_i$ vary slowly everywhere in space, one can consider $\vb{u}_i \mapsto \vb{u}(\vb{x})$ as a smooth field and expand expand the elastic potential $V[\vb{r}]$ to quadratic order in the gradients $\partial_{\alpha} u_{\beta}(\vb{x})$ ($\alpha, \beta = x,y$), leading to the following standard elastic Hamiltonian \cite{chaikinPrinciplesCondensedMatter2013}:
\begin{equation}
\label{eq:H_el}
	\begin{split}
		H_{\el}[\vb{u}(\vb{x})] &= \frac{1}{2} \sum_{ij,\alpha \beta} \pdv[2]{V}{u_{\alpha,i}}{u_{\beta,j}} u_{\alpha,i} u_{\beta,j} \\
		&\simeq \frac{1}{2} \int \dd[2]{\vb{x}} C^{\alpha \beta \gamma \delta} u_{\alpha \beta} u_{\gamma \delta} ,
	\end{split}
\end{equation}
where $u_{\alpha \beta} \equiv \frac{1}{2} (\partial_{\alpha} u_{\beta} + \partial_{\beta} u_{\alpha})$ is called the symmetric strain tensor, and $C^{\alpha \beta \gamma \delta}$ is a tensor of elastic constants. The long-wavelength normal modes of $H_{\el}$, called phonons, are gapless Goldstone modes associated with the broken translational symmetry. Using the Gaussian Hamiltonian \eqref{eq:H_el}, one can immediately compute the low-temperature fluctuations of the translational order parameters. Similar to the case of ferromagnetic order in the two-dimensional XY model, one finds that long-range translational order is replaced by quasi-long-range order (QLRO) at low nonzero temperatures \cite{chaikinPrinciplesCondensedMatter2013}, as evinced by the correlation function:
\begin{equation}
\label{eq:density_corr}
	\expval{e^{i \vb{G} \cdot (\vb{r}_i - \vb{r}_j)}} \simeq \expval{e^{i \vb{G} \cdot [\vb{u}(\vb{x}) - \vb{u}(0)]}} \sim \frac{1}{\abs{\vb{x}}^{\eta_{\vb{G}}}} ,
\end{equation}
where $\vb{x}$ is the equilibrium displacement between atoms $i$ and $j$, and the nonuniversal exponent $\eta_{\vb{G}}$ depends on the magnitude of $\vb{G}$ and the elastic constants $C^{\alpha \beta \gamma \delta}$.

\textit{(2) Domain walls}: for sufficiently large $C^{\alpha \beta \gamma \delta}$, when atoms are arranged into an approximate square lattice, we can regard each Ising spin as effectively interacting with only its four nearest neighbors. Thermal fluctuations in the spin degrees of freedom are then effectively described by an ordinary AF Ising model on the square lattice, with the Hamiltonian
\begin{equation}
\label{eq:H_AF}
	H_{\AF}[\sigma] = J \sum_{\expval{ij}} \sigma_i \sigma_j .
\end{equation}
Flipping a domain of spins results in a domain wall at its boundary, which costs energy $2J$ per unit length (in units of the temperature). As in an ordinary Ising model, these domain walls proliferate below a critical coupling strength $J_c$, resulting in a paramagnetic (PM) solid\footnote{We assume, for simplicity, that the presence of AF interactions is sufficient to stabilize the square lattice even in the paramagnetic phase; i.e., we do not consider the possibility of a structural phase transition to a triangular lattice, which would exhibit highly frustrated antiferromagnetism.}. Notably, the staggered magnetization \eqref{eq:stag_mag} does not exhibit long-range order at any nonzero temperature, owing to the lack of long-range translational order. Instead, staggered spin-spin correlation functions transition from QLRO above $J_c$ to short-range order below $J_c$:
\begin{equation}
\label{eq:staggered_corr}
	\expval{\sigma_i \sigma_j e^{-i \vb{Q} \cdot (\vb{r}_i - \vb{r}_j) }} \sim \begin{dcases}
		m^2 |\vb{x}|^{-\eta_{\vb{Q}}}, & J > J_c \\
		e^{-\abs{\vb{x}}/\xi} , & J < J_c
	\end{dcases} ,
\end{equation}
where we have neglected a subleading power-law prefactor in the $J < J_c$ case.

\begin{figure}[t]
	\centering
	\includegraphics[width=\columnwidth]{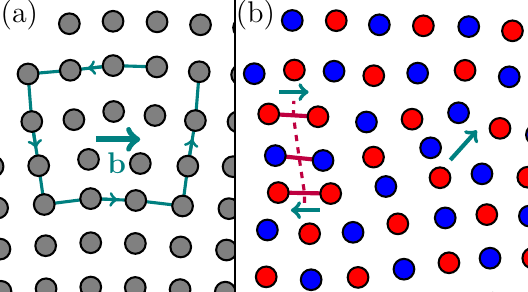}
	\caption{(a) An elementary dislocation of Burgers vector $\vb{b} = a \hat{\vb{x}}$ in a square lattice. While the atoms are locally arranged in a regular square lattice away from the dislocation core, a semi-infinite ``missing'' line of atoms terminates at the dislocation core. As a result, a Burgers circuit traversed around the dislocation will fail to close. The resulting slow modulation of the lattice placement in space leads to logarithmic interactions between far-away dislocations. (b) Left: two elementary dislocations in the presence of strong AF interactions between atoms; red (blue) atoms denote $\sigma_i = +1$ ($\sigma_i = -1)$. Since elementary dislocations disrupt the bipartiteness of the lattice, they are necessarily attached to the endpoints of string-like domain walls. As a result, elementary dislocations are \textit{linearly} confined in the presence of strong AF interactions. Right: a double dislocation with Burgers vector $\vb{b} = a (\hat{\vb{x}} + \hat{\vb{y}})$ does not disrupt the bipartiteness of the lattice, and does not have any long-range effect on the AF order as a result.}
	\label{fig:dislocations}
\end{figure}
	
\textit{(3) Dislocations}: analogous to vortices in the XY model, two-dimensional solids host pointlike topological excitations called dislocations which cannot be constructed from smooth phonon excitations \cite{chaikinPrinciplesCondensedMatter2013}. Microscopically, a dislocation can be visualized in a square lattice by removing a semi-infinite line of atoms and stitching the solid back together, as in Fig.~\ref{fig:dislocations}(a). Just above the line of missing atoms is the dislocation core, where the periodic arrangement of atoms breaks down. Far away from the vortex core, perfect crystalline order is maintained \textit{locally}; however, the missing line of atoms results in a slow ``winding'' of the lattice placement in space. The dislocation can be detected by traversing a Burger's circuit about the dislocation core: upon adding the local relative displacements $\partial_{\alpha} \vb{u}$ along a closed contour $\gamma$ encircling the dislocation, one finds that the displacements do not add to zero as in the absence of a dislocation. Instead, the missing line of atoms results in a total displacement $\vb{b}$ called the Burgers vector, which is necessarily a lattice vector:
\begin{equation}
	\oint_{\gamma} \dd{\ell^{\alpha}} \partial_{\alpha} \vb{u} = \vb{b} = a(n \hat{\vb{x}} + m \hat{\vb{y}}) \quad (n, m \in \mathbb{Z}) .
\end{equation}
Due to the elastic rigidity of the lattice, the slowly winding translational order results in a large logarithmic energy cost of a single dislocation; as a result, finite energy configurations must have zero total Burgers vector. The total elastic energy in the presence of dislocations is given by\footnote{Strictly speaking, the form of $G^{\text{dis}}_{\alpha \beta}(\vb{x})$ in Eq.~\eqref{eq:dislocInt} is appropriate only for the case in which $C^{\alpha \beta \gamma \delta}$ has only the two independent elastic constants of the \textit{triangular} lattice. The square lattice admits a third independent elastic constant, which can introduce an additional anisotropy into the interactions between dislocations, but does not modify the leading logarithmic growth of the interactions.}
\begin{equation}
	\begin{split}
		H_{\el}[\vb{u}(\vb{x})] &= H_{\el}[\tilde{\vb{u}}(\vb{x})] - \frac{1}{2} \sum_{n \neq m} b^{\alpha}_n G^{\text{dis}}_{\alpha \beta}(\vb{x}_n - \vb{x}_m) b^{\beta}_m , \\
		G^{\text{dis}}_{\alpha \beta}(\vb{x}) &= \frac{C}{2\pi} \qty( \delta_{\alpha \beta} \log \frac{\abs{\vb{x}}}{a} - \frac{x_{\alpha} x_{\beta}}{\vb{x}^2} ) ,
	\end{split}
 \label{eq:dislocInt}
\end{equation}
where $\tilde{\vb{u}}(\vb{x})$ is the smooth part of $\vb{u}(\vb{x})$ (i.e., the phonon contribution with no dislocations), while the constant $C$ is a function of the various eastic constants $C^{\alpha \beta \gamma \delta}$. We see that pairs of dislocations with oppositely-signed Burgers vectors are logarithmically bound to each other at low temperatures. 

In an ordinary two-dimensional crystal without AF interactions, the proliferation of dislocations below a critical elastic coupling strength $C$ destroys the solid's translational QLRO; i.e., density-density correlation functions of the form \eqref{eq:density_corr} decay exponentially below a critical value of $C$. Interestingly, the proliferation of dislocations do not entirely destroy the solid's orientational order: instead, the resulting phase is called a \textit{tetratic}\footnote{In the more familiar case of a triangular lattice, the phase obtained from proliferating dislocations would be called a \textit{hexatic}. The naming reflects the residual rotational symmetry of the phase: the tetratic has QLRO in an order parameter field that is invariant under $\frac{\pi}{2}$ rotations, while the hexatic has QLRO in an order parameter field invariant under $\frac{\pi}{3}$ rotations.} phase, and is characterized by translational short-range order and orientational QLRO. The destruction of orientational order proceeds via the proliferation of another class of topological defects called \textit{disclinations}, resulting in a liquid phase. For simplicity, we shall focus on the solid and tetratic phases in this work and neglect disclinations entirely.

The preceding discussion of dislocations applies in the absence of AF interactions. In the presence of strong antiferromagnetism, a new feature emerges: since dislocations disrupt the bipartiteness of the lattice, they necessarily frustrate the AF order. As shown in Fig.~\ref{fig:dislocations}(b), an elementary dislocation/anti-dislocation pair separated by a distance $R$ is necessarily connected by a string-like Ising domain wall with minimum energy cost $2JR$, resulting in a \textit{linear} confinement of elementary dislocations. In contrast, double dislocations do not disrupt the bipartiteness of the lattice and experience only a logarithmic confinement. Ref.~\cite{abutbulTopologicalOrderAntiferromagnetic2022} therefore suggested the possibility of an AF tetratic phase in which double dislocations with Burgers vectors $\abs{\vb{b}} = a \sqrt{2}$ proliferate while elementary dislocations remain confined.

\begin{figure}[t]
	\centering
	\includegraphics[width=0.85\columnwidth]{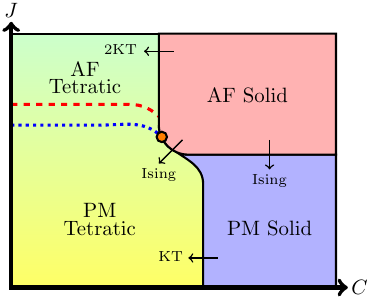}
	\caption{Schematic phase diagram of the classical antiferromagnetic melting problem in two spatial dimensions, focusing only on the solid/tetratic phases. Arrows between phases indicate the expected universality classes of each transition; $2KT$ refers to a Kosterlitz-Thouless transition in which only \textit{double} dislocations proliferate. As we show in this work, the AF tetratic and PM tetratic regimes are \textit{thermodynamically} the same phase, but can be distinguished by a \textit{computational} phase transition which attempts to bipartition the atoms into two sublattices. The red dotted line denotes the algorithm-dependent transition arising from a given computational protocol, while the blue dotted line denotes an ``intrinsic'' computational phase transition below which \textit{no} algorithm can consistently bipartition the atoms with unit probability. The orange dot indicates an expected multicritical point, where the 2KT and Ising transitions meet.}
	\label{fig:phase_diagram}
\end{figure}

We are thus led to the schematic phase diagram presented in Fig.~\ref{fig:phase_diagram}. Starting from large $J$ and $C$, the two-step transition from AF solid, to PM solid, and finally to PM tetratic, proceeds via relatively conventional thermodynamic phase transitions; we expect these transitions to lie in the Ising and Kosterlitz-Thouless (KT) universality classes respectively, although the coupling of the Ising spins to gapless phonons can in principle modify the universality class of the former transition. One novel feature of the phase diagram is a potential direct transition from the AF solid to the PM tetratic in the Ising universality class. The possibility of such a transition was first noted in the present context in Ref.~\cite{cardyTransitionsInternalDegree1983}, and is closely related to an analogous Ising phase transition in a modified XY model \cite{shiBosonPairingUnusual2011}, as elaborated in the discussion. This Ising transition is expected to meet the 2KT transition, where double dislocations proliferate while single dislocations remain confined, at a multicritical point \cite{sernaDeconfinementTransitionsGeneralised2017} (indicated by an orange dot).

An outstanding question remains on the nature of the transition between the putative AF and PM tetratic phases. Importantly, staggered spin-spin correlation functions of the form (\ref{eq:staggered_corr}) decay exponentially \textit{regardless} of the nature of the Ising ordering, simply due to positional disorder. As a result, it is a nontrivial task to construct an order parameter which can distinguish between the AF and PM tetratic phases. In the remainder of this work, we shall present two distinct perspectives on the nature of this transition. First, we demonstrate that the AF and PM tetratic regimes are \textit{thermodynamically} the same phase; in particular, we shall argue that no ordinary \textit{local} order parameter, or correlations between local observables, can detect a sharp phase transition between these regimes. Next, we demonstrate that a \textit{nonlocal} AF order parameter, which shall generally require a nontrivial classical computation to define, can be constructed to distinguish these regimes. This order parameter then undergoes a non-thermodynamic \textit{computational} transition as the AF interaction strength is reduced.

\section{Absence of Thermodynamic AF/PM Tetratic Phase Transition}
\label{sec:noPT}

In this section, we shall demonstrate that the AF and PM tetratic phases are adiabatically connected to each other. At first sight, if AF and positional orders are regarded as distinct order parameters corresponding to independent symmetries, then nothing would prevent the AF/PM and solid/tetratic phase transitions from occurring independently in either order. However, it must be emphasized that the AF and positional orders are \textit{not} independent. This can be clearly seen from simple Ginzburg-Landau theory considerations: since $M_{\vb{Q}}$ transforms nontrivially under uniform translations, $\expval{M_{\vb{Q}}} \neq 0$ indicates both Ising \textit{and} translational symmetry-breaking\footnote{For another perspective, if $M_{\vb{q}}$ and $\rho_{\vb{q}}$ represent the Fourier components of the spin density and the atomic density respectively, then the Landau free energy will generically contain terms such as $\sum_{\vb{q}} (\rho_{2\vb{q}}^* [M_{\vb{q}}]^2 + \text{c.c.})$, which induce a density wave of wavelength $\lambda$ whenever a spin-density wave of wavelength $2\lambda$ is present.}. Here we will present a complementary explanation for the lack of (quasi)-long range AF order within the nominal AF tetratic phase, which will assist in identifying an algorithm which can computationally construct an AF order parameter. 

The central observation is that, in the presence of AF order, elementary dislocations carry Ising gauge flux. To illustrate this point, consider the pair of dislocations in Fig.~\ref{fig:dislocations}(b). Far away from the dislocation cores, the atoms are arranged in a semi-regular square lattice and the Ising interactions can be modeled using the nearest-neighbor Hamiltonian \eqref{eq:H_AF}. In the presence of dislocations, it is conceptually useful to trivially rewrite this Hamiltonian as
\begin{equation}
	H_{\AF}[\sigma] = -J \sum_{\expval{ij}} \sigma_i U_{ij} \sigma_j, \quad U_{ij} = -1 .
\end{equation}
The numbers $U_{ij} = -1$, defined on each nearest-neighbor link $\expval{ij}$ of the lattice, can be considered as a fixed background $\mathbb{Z}_2$ gauge field in a particular gauge. In a perfect square lattice, such a gauge field would be ``flat", i.e., gauge-equivalent to the trivial gauge field\footnote{Note, however, that if a periodic lattice contains an odd number of rows or columns, then the $U_{ij} = -1$ gauge field is no longer trivial, since the Wilson loop $W_{\gamma} = -1$ for paths around the non-contractible loops of the torus. In other words, an Ising antiferromagnet on a torus with an odd number of rows or columns is equivalent to an Ising ferromagnet with a $\mathbb{Z}_2$ gauge flux through one or both of the holes of the torus.} in which $U_{ij} = +1$ everywhere. However, this is no longer the case on a lattice with dislocations. Indeed, if $\gamma$ denotes a closed path through the lattice, then $\gamma$ must traverse an odd number of links whenever it encircles a single elementary dislocation. As a result, the ``Wilson loop'' constructed from the gauge field $U_{ij}$ on such a path yields
\begin{equation}
    \prod_{\expval{ij} \in \gamma} U_{ij} = -1 .
\end{equation}
This equation simply states that the loop $\gamma$ necessarily crosses an odd number of domain walls; i.e., the endpoint of an open-string domain wall necessarily terminates on the dislocation. Notably, this holds even for loops far away from the dislocation core where the local lattice structure is largely unaffected by the dislocation. 

If we assume that the positional correlation length (roughly, the typical distance between double dislocations) is much larger than the lattice spacing, then the essential interplay between Ising and positional degrees of freedom is captured by the gauge flux of the dislocations alone. Thus, a physically reasonable model for the Ising spins {\it within the tetratic phase} is a ``gauged'' $\mathbb{Z}_2$ Ising model on a \textit{regular} square lattice, sometimes known as the two-dimensional Fradkin-Shenker (FS) model \cite{fradkinPhaseDiagramsLattice1979}:
\begin{equation}
\label{eq:H_FS}
	H_{\FS} [\sigma, U] = -J \sum_{\expval{ij}} \sigma_i U_{ij} \sigma_j - g \sum_{[ijk\ell]} U_{ij} U_{jk} U_{k\ell} U_{\ell i} ,
\end{equation}
where $U_{ij}$ is now a \textit{dynamical} $\mathbb{Z}_2$ gauge field, defined on the links of a regular square lattice. The latter sum over all plaquettes $[ijk\ell]$ with vertices $i,j,k,\ell$ provides a fugacity $e^{-2g}$ to each Ising flux, i.e., plaquettes $[ijk\ell]$ on which $U_{ij} U_{jk} U_{k\ell} U_{\ell i} = -1$, which model the elementary dislocations in the tetratic phase. Just as open-string domain walls terminate on dislocations in the AF tetratic, open-string domain walls terminate on Ising fluxes in the FS model. Note that this model only applies to the tetratic phase because it neglects the logarithmic interactions that act between dislocations in the solid phase; if desired, these interactions can be added to analyze the full thermodynamic phase diagram \cite{cardyTransitionsInternalDegree1983}.

It is well-known that the Hamiltonian \eqref{eq:H_FS} has a trivial thermodynamic phase diagram in two dimensions \cite{fradkinPhaseDiagramsLattice1979}. Naively, one might expect the large $J$, $g$ limit to correspond to a Higgs phase in which fluxes are linearly confined, while the small $J$, $g$ limit would correspond to a confined phase in which fluxes have proliferated; the former would correspond to the AF tetratic phase of the model \eqref{eq:ham_schematic} while the latter would correspond to the PM tetratic phase. However, these two phases are in fact separated by a smooth crossover, rather than a sharp phase transition. A physically transparent method of verifying this is to notice that the Hamiltonian \eqref{eq:H_FS} is Kramers-Wannier dual to an Ising model in a symmetry-breaking field $h$ \cite{kardarStatisticalPhysicsFields2007,nussinovDerivationFradkinShenkerResult2005}. The strength of $h$ in the dual Ising model is related to the flux fugacity via $\tanh h = e^{-2g}$; thus, the crossover approaches a true phase transition only in the limit $g \to \infty$ where fluxes are forbidden, in which case (\ref{eq:H_FS}) is equivalent to an ordinary Ising model.

Finally, we discuss the fate of staggered spin-spin correlations in the presence of dislocations. With the FS model, the closest analog to these spin-spin correlations are the string correlation functions $G_{\gamma}(i,j)$ on a path $\gamma$ connecting the lattice sites $i$ and $j$:
\begin{equation}
\label{eq:FS_string_corr}
	G_{\gamma}(i,j) = \expval{ \sigma_i \qty[ \prod_{\expval{k\ell} \in \gamma} U_{k\ell} ] \sigma_j } .
\end{equation}
In the case $g \to \infty$ where Ising fluxes are absent, $G_{\gamma}$ depends only on the endpoints $i$ and $j$ and exhibits long-range order for $J > J_c$. Indeed, if one picks the gauge $U_{ij} = -1$ everywhere, $G_{\gamma}$ reduces exactly to the staggered spin-spin correlation function in a square-lattice Ising antiferromagnet. In contrast, for any $g < \infty$, $G_{\gamma}$ decays exponentially with the length of the string $\gamma$. In a low-temperature expansion (i.e., for $J$ and $g$ large), this decay occurs due to a dilute gas of tightly-bound flux pairs localized to the string $\gamma$, each of which flips the sign of the product $\prod_{\expval{k\ell} \in \gamma} U_{k\ell}$ \cite{fradkinPhaseDiagramsLattice1979}. 

In the language of the original AF tetratic, $G_{\gamma}$ defines staggered correlation functions between spins $\sigma_i$ and $\sigma_j$ by replacing the factor $e^{-i \vb{Q} \cdot (\vb{r}_i - \vb{r}_j)}$ in Eq.~\eqref{eq:staggered_corr} with the factor
\begin{equation}
    (-1)^{|\gamma|} \equiv \exp{i \vb{Q} \cdot \int_{\gamma} \dd{\ell}^{\alpha} \partial_{\alpha} \vb{u}} ,
\end{equation}
where $\gamma$ is a \textit{fixed} path through the lattice connecting atoms $i$ and $j$ and $|\gamma|$ is its length. The exponential decay of the resulting correlation function $\expval{(-1)^{|\gamma|} \sigma_i \sigma_j}$ then arises due to dislocation pairs which cross the path $\gamma$, changing the length of $\gamma$ by one unit. Notably, this exponential decay occurs even in the AF solid phase, so long as elementary dislocations have any nonzero fugacity. In this language, it is clear that such a correlation function does not properly characterize the strength of AF correlations actually observed for large $J$. Instead, as we shall elaborate in the following, one can allow for the path $\gamma$ to be chosen \textit{dynamically}, conditioned on the locations of dislocations, so as to avoid tightly-bound dislocation pairs in an otherwise strongly-ordered antiferromagnet.

\section{Computational Bipartiteness Transition}
\label{sec:computational}

We now turn to the main results of this work, which demonstrate that the AF and PM tetratic regimes can be distinguished by a \textit{computational} phase transition. In Sec.~\ref{subsec:comp_overview} we qualitatively explain how an AF order parameter can be \textit{algorithmically} constructed within the tetratic phase, whose value is large deep in the AF tetratic regime and is expected to vanish in a continuous computational phase transition as the AF interactions are weakened. While the strength and transition point of such an order parameter generally depends on the precise algorithm by which it is constructed, an algorithm-independent distinction can be made between the AF and PM tetratic regimes by employing periodic boundary conditions: namely, in the AF tetratic regime it is always possible to construct a well-defined bipartitioning of the atoms into two sublattices, while in the PM tetratic regime the atoms cannot be consistently bipartitioned when topologically nontrivial boundary conditions are imposed.  

In Secs.~\ref{subsec:fs_mwpm}, \ref{subsec:fs_bipartiteness} and \ref{subsec:optimal}, we demonstrate these ideas explicitly in the analytically and numerically tractable context of the Fradkin-Shenker (FS) model. In Sec.~\ref{subsec:fs_mwpm}, we show analytically and numerically that a simple minimal-weight pairing algorithm can pair gauge fluxes to establish a \textit{ferromagnetic} order parameter, which then undergoes a computational phase transition separating ferromagnetic and paramagnetic computational phases. In Sec.~\ref{subsec:fs_bipartiteness} we show that the FS model undergoes a ``bipartiteness transition'' in  the presence of topologically nontrivial boundary conditions: in the bipartiteable phase the homology class of the domain walls can be inferred with high probability from the positions of the gauge fluxes, while in the non-bipartiteable phase it cannot. As a result, \textit{any} flux-pairing algorithm will frequently fail to construct a well-defined computational order parameter in the non-bipartiteable phase. Finally, in Sec.~\ref{subsec:optimal} we construct an algorithm for pairing gauge fluxes which establishes a nonzero computational order parameter throughout the entire bipartiteable phase; we conjecture that this algorithm is ``optimal'', in the sense that it establishes a nonzero order parameter in the largest possible parameter regime.

\begin{figure*}[t]
	\centering
	\includegraphics[width=\textwidth]{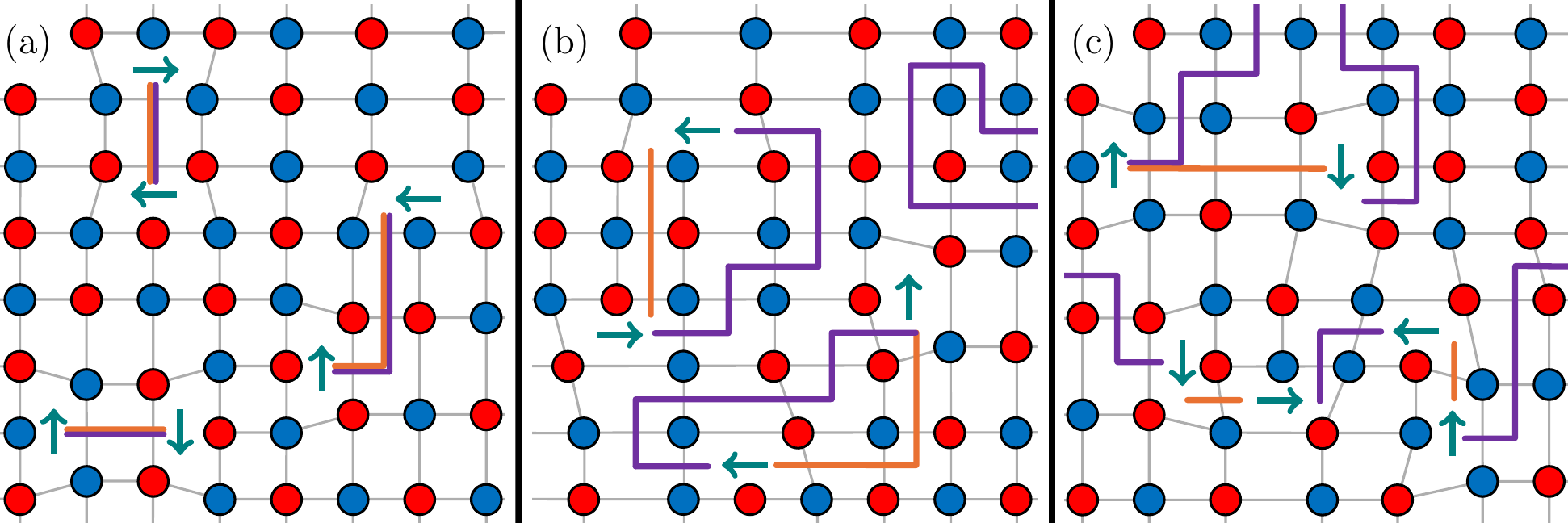}
	\caption{Schematic depiction of our dislocation pairing protocol within the tetratic phase. By pairing dislocations, a bipartitioning of the atoms into sublattices $A$ and $B$ is established, allowing for the computation of a staggered magnetization and correlation functions [Eqs.~\eqref{eq:stag_mag_2} and \eqref{eq:stag_corr_2}, respectively]. Blue/red circles denote up/down spins, green arrows denote dislocations, purple lines denote physical domain walls, and orange lines denote pairings of dislocations. (a) When $J$ is very large, the graph remains ``almost bipartite": aside from a dilute density of tightly-bound dislocation pairs, the spins appear to exhibit strong AF order. Once well-defined domains are established by pairing dislocations, the majority of spins belong to the same AF domain. (b) As $J$ is reduced and domain walls begin to fluctuate, although it becomes increasingly difficult to visually pair dislocations, a classical algorithm can easily establish a minimal-weight pairing. The physical domain walls and pairing paths together form the boundaries of AF domains. The computationally constructed staggered magnetization $M_{\stag}$ remains nonzero so long as these domains do not percolate. (c) In a system with periodic boundary conditions, it is possible to pair dislocations in such a way that the resulting lattice is not bipartite and $M_{\stag}$ remains ill-defined. This occurs when the pairing paths and physical domain walls together form a non-contractible cycle around the torus. If the homology class of the domain walls cannot be determined with high probability from the positions of dislocations, then \textit{no} dislocation-pairing protocol can consistently establish an AF order parameter.}
	\label{fig:pairing_protocol}
\end{figure*}

\subsection{Qualitative Overview}
\label{subsec:comp_overview}
Fundamentally, the absence of a transition between the AF and PM tetratic regimes can be ascribed in part to the lack of a well-defined AF order parameter. Deep in the solid phase where dislocations are absent, the lattice has a well-defined \textit{bipartite structure}, i.e., atoms can be divided into two sublattices $A$ and $B$ such that the AF interactions only occur significantly between atoms of opposite sublattices. A staggered magnetization $M_{\stag}$ can then be defined simply by subtracting the magnetizations of opposite sublattices:
\begin{equation}
\label{eq:stag_mag_2}
	M_{\stag} \equiv M_A - M_B = \frac{1}{N} \qty( \sum_{i \in A} \sigma_i - \sum_{i \in B} \sigma_i ) .
\end{equation}
Similarly, staggered spin-spin correlation functions can be immediately defined by introducing signs $s_i = \pm 1$ which indicate the sublattice of each site:
\begin{equation}
\label{eq:stag_corr_2}
	G_{\stag}(i,j) = s_i s_j \expval{\sigma_i \sigma_j}, \quad s_{i} = \begin{cases}
		+1, & i \in A \\
		-1, & i \in B
	\end{cases} .
\end{equation}
These definitions of the staggered magnetization and correlation function agree with that of Eq.~\eqref{eq:stag_mag} (up to a possible overall phase) and Eq.~\eqref{eq:staggered_corr} respectively when the atomic positions are locked in a perfect square lattice. Importantly, they remain well-defined once the atomic positions are allowed to fluctuate, so long as no \textit{elementary} dislocations are present; as noted by Ref.~\cite{abutbulTopologicalOrderAntiferromagnetic2022}, double dislocations preserve the bipartiteness of the lattice. Thus, in the limit that elementary dislocations are completely forbidden, $M_{\stag}$ remains both well-defined and nonzero across the AF solid/tetratic transition, and $G_{\stag}$ remains long-range ordered. 

As $J$ is reduced and elementary dislocations are introduced, the staggered magnetization \eqref{eq:stag_mag_2} and staggered correlation function \eqref{eq:stag_corr_2} immediately become problematic: strictly speaking, a lattice with any number of elementary dislocations is no longer bipartite, and it is not a priori clear how to consistently sort atoms into $A$ and $B$ sublattices. However, if $J$ is sufficiently large, dislocations remain strongly confined by string-like domain walls, and the lattice remains ``almost bipartite''. We can then naturally bipartition the atoms by ``pairing'' dislocations, i.e., drawing paths through the dual lattice which connect the dislocations pairwise [see Fig.~\ref{fig:pairing_protocol}(a)]. Nearest-neighbor pairs of atoms away from these pairing paths are sorted into opposing sublattices as usual, while nearest-neighbor pairs which are bisected by these pairing paths are sorted into the same sublattice.

As $J$ is reduced further, and dislocations become more weakly confined, we require a systematic procedure for pairing dislocations. In order to construct physically meaningful AF observables from such a dislocation-pairing protocol, it is crucial that each dislocation pairing is decided based on the atomic positions $\vb{r}$ alone, and is agnostic to the values of the Ising spins. One natural algorithm we can employ is \textit{minimal-weight pairing}. In the first step, one assigns each atom a set of nearest neighbors to construct a lattice and identify dislocations in the resulting lattice; we describe one concrete protocol for performing this step in Sec.~\ref{sec:hardspheres}. In the second step, we choose a pairing of the elementary dislocations by demanding that the pairing path bisects the fewest possible nearest-neighbor pairs. Such an optimization problem can be cast as a problem in integer linear programming which can be solved in polynomial time \cite{edmondsPathsTreesFlowers1965,barahonaMorphologyGroundStates1982}. When AF correlations are strong and domain walls are costly, these pairing paths will naturally coincide with the trajectories of domain walls connecting dislocations, and this minimal-weight pairing will sort all atoms into the same effective domain. As $J$ is reduced the domain walls will begin to fluctuate away from these minimal-weight trajectories, resulting in flipped domains of spins as in Fig.~\ref{fig:pairing_protocol}(b).

In an infinite system, or a system with open boundary conditions, this minimal-weight pairing prescription provides a means to compute observables such as the staggered magnetization \eqref{eq:stag_mag_2} and correlation function \eqref{eq:stag_corr_2}. It is clear that for large $J$, minimal-weight pairing will establish a nonzero $M_{\stag}$ and long-range ordered correlations $G_{\stag}$; in contrast, for sufficiently small $J$, no pairing will result in long-range AF order. Thus, we expect that the computationally defined observables $M_{\stag}$ and $G_{\stag}$ will undergo an order/disorder phase transition as $J$ is tuned. We emphasize once again that this \textit{computational} phase transition does not indicate a \textit{thermodynamic} transition, since $M_{\stag}$ and $G_{\stag}$ are highly nonlocal observables which must be defined via a nontrivial classical computation.

In a system with periodic boundary conditions, our dislocation-pairing protocol can exhibit a particularly interesting mode of failure: specifically, it is possible to pair dislocations in such a way that the resulting lattice is not bipartite. This occurs whenever bonds cut by the dislocation pairing and bonds cut by domain walls together form an odd number of non-contractible cycles around a given hole of the torus, as in Fig.~\ref{fig:pairing_protocol}(c). In such an event, one can traverse a closed path through the system and switch sublattices an odd number of times, indicating that the atoms have not been consistently bipartitioned. 

For large $J$, the domain walls corresponding to a particular configuration of dislocations are statistically guaranteed to fall into a specific \textit{homology class}\footnote{Two paths through the dual lattice with the same endpoints are called \textit{homologous} if their union bounds a closed region of sites. A \textit{homology class} is an equivalence class of homologous paths. In a system with open boundary conditions, any two paths with the same endpoints are homologous; i.e., there is only one homology class for each set of endpoints.}; if this homology class can be efficiently determined by a classical computation, then we can always choose a pairing of dislocations which falls into the same homology class, resulting in a bipartite lattice. In contrast, if $J$ is sufficiently small that domain walls of differing homology classes exhibit comparable probabilities, then with non-vanishing probability \textit{any} method of pairing dislocations will fail to establish a bipartitioning. As we shall see, the difference between these two cases is again distinguished by a sharp computational phase transition.

We can therefore attempt to sharply distinguish between AF and PM tetratic regimes by the following computational protocol. First, we sample a configuration $\qty{\vb{r}_i, \sigma_i}$ of the atoms, using periodic boundary conditions in both spatial directions. Then, assuming that the positional correlation length is appreciably larger than the lattice spacing, we determine the lattice structure and position of dislocations. Next, using the locations of the dislocations alone, we attempt to pair dislocations in order to construct a bipartite lattice. If the ``correct'' pairing can be determined with unit probability in the thermodynamic limit, we say that we are in the AF tetratic regime; otherwise, we say that we are in the PM tetratic regime. As we elaborate in the discussion (Sec.~\ref{sec:discussion}), this protocol is analogous to that of optimal quantum error correction protocols in surface codes.

\subsection{Minimal-Weight Pairing in the Fradkin-Shenker Model}
\label{subsec:fs_mwpm}

It is possible, although technically challenging, to numerically implement the proposed dislocation-pairing protocol and observe the previously described computational phase transitions in microscopically realistic models of AF melting, such as in Refs.~\cite{abutbulTopologicalOrderAntiferromagnetic2022,jalal_thesis} (see also Sec.~\ref{sec:hardspheres} below). Instead, in order to demonstrate the dislocation-pairing idea in the simplest possible setting, we will consider the analogous computational phase transitions in the FS model defined by the Hamiltonian in Eq.~\eqref{eq:H_FS} on a regular square lattice. Specifically, we will show in this section how a classical algorithm can be used to construct nonlocal \textit{ferromagnetic} observables in the FS model which exhibit an order/disorder phase transition tuned by the Ising coupling $J$, despite the absence of any such thermodynamic transition. Throughout this section we shall be somewhat cavalier about boundary conditions; as mentioned in the previous section and to be elaborated in Sec.~\ref{subsec:fs_bipartiteness}, the algorithm we describe can potentially fail to construct consistent observables in the presence of topologically nontrivial boundary conditions.

In what follows, it shall prove useful to eliminate the gauge redundancy from the FS model and work with manifestly gauge-invariant degrees of freedom. The full physical content of the FS model is contained in the gauge-invariant domain wall variables $V_{ij} = \sigma_i U_{ij} \sigma_j$, in terms of which the Hamiltonian can be rewritten as
\begin{equation}
\label{eq:H_FS_2}
    H_{\FS}[V] = -J \sum_{\expval{ij}} V_{ij} - g \sum_{[ijk\ell]} V_{ij} V_{jk} V_{k\ell} V_{\ell i} .
\end{equation}
Similarly, string correlators $G_{\gamma}$ defined in Eq.~\eqref{eq:FS_string_corr} can be expressed in terms of $V_{ij}$ as
\begin{equation}
    G_{\gamma} = \expval{ \prod_{\expval{ij} \in \gamma} V_{ij} } .
\end{equation}
We denote by $V = \qty{V_{ij}}$ an arbitrary domain wall configuration, with domain walls graphically corresponding to paths through the dual lattice along which $V_{ij} = -1$. Ising fluxes, i.e., plaquettes $[ijk\ell]$ where $V_{ij} V_{jk} V_{k\ell} V_{\ell i} = -1$, graphically correspond to endpoints of these paths. The product $\prod_{\expval{ij} \in \gamma} V_{ij}$ then counts the number of times (mod 2) that an open path $\gamma$ in the direct lattice crosses a domain wall.

As discussed in Sec.~\ref{sec:noPT}, dislocations in the AF tetratic are modeled within the FS model as Ising fluxes. In the limit $g \to \infty$ where fluxes are forbidden, domain walls must form closed loops in the dual lattice, resulting in well-defined domains of aligned spins. Consequently, the string correlators $G_{\gamma}$ depend only on the string's initial and final endpoints and are otherwise path-independent. In the gauge $U_{ij} = +1$ everywhere, these string correlators reduce to ordinary ferromagnetic Ising correlation functions $\expval{\sigma_i \sigma_j}$ and exhibit long-range ferromagnetic order for large $J$. This ferromagnetic order in the FS model in the absence of fluxes is analogous to AF order in the tetratic phase in the absence of any elementary dislocations.

In contrast, for $g < \infty$, the FS model admits both ordinary closed-loop domain walls and open-string domain walls. As a result, the physical spins cannot a priori be divided into well-defined domains, and one cannot establish an Ising order parameter. Additionally, the product $\prod_{ij \in \gamma} V_{ij}$ becomes path-dependent, changing sign each time the path is deformed through a flux, and $G_{\gamma}$ decays exponentially in the length of the string $\gamma$ due to tightly-bound flux pairs which straddle the path $\gamma$.

\begin{figure*}[t]
\includegraphics[width=\textwidth]{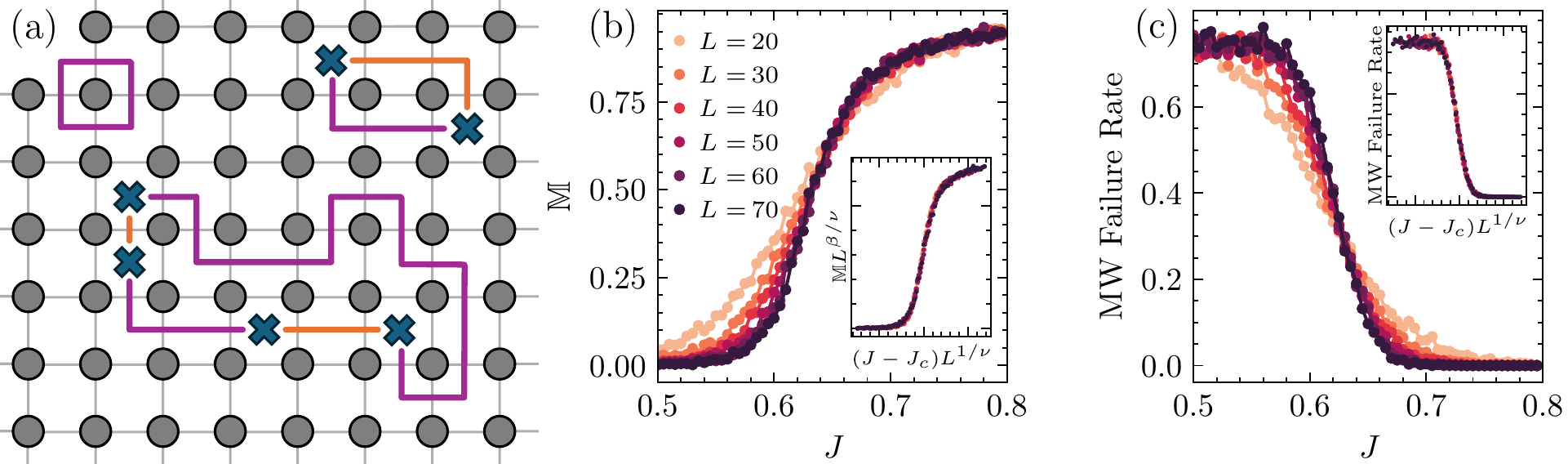}
\caption{(a) Minimal-weight pairing of Ising fluxes in the Fradkin-Shenker model. Fluxes are represented by blue crosses, domain walls by purple lines, and pairing paths by orange lines. Once a pairing is established, the domain walls and pairing paths together define domains of aligned spins. (b,c) Two computational observables in the Fradkin-Shenker model [Eq.~\eqref{eq:H_FS}] with $g = 1$, computed numerically using Monte-Carlo simulations and minimal-weight pairing of Ising fluxes, for square $L \times L$ periodic lattices of various linear system sizes $L$. Each data point is averaged over 1000 realizations. (b) Computational magnetization $\mathbb{M}$ [Eq.~\eqref{eq:FS_M_comp}] established via minimal-weight pairing. A computational ``ferromagnetic'' regime with nonzero $\mathbb{M}$ at large $J$, and a computational ``paramagnetic'' regime with vanishing $\mathbb{M}$ at small $J$, are separated by a (non-thermodynamic) continuous computational phase transition. Inset: finite-size scaling collapse using the estimated critical point $J_{c,\MW} \simeq 0.6298$ and the estimated critical exponents $\nu_{\MW} \simeq 1.4015$ and $\beta_{\MW} \simeq 0.1096$. (c) In the presence of topologically nontrivial boundary conditions, minimal-weight pairing can fail to consistently establish a computational magnetization. In the thermodynamic limit, this occurs with vanishing probability in the computational ferromagnetic regime, and with $75\%$ probability in the computational paramagnetic regime (for periodic boundary conditions in both directions). Inset: finite-size scaling collapse using the same estimated critical point $J_{c,\MW}$ and critical exponent $\nu_{\MW}$ as in (b).}
\label{fig:mwpm_numerics}
\end{figure*}

While string correlations for a \textit{fixed} path $\gamma$ decay exponentially in the presence of fluxes, there is nevertheless a sense in which typical microstates of the system appear strongly ferromagnetic for large $J$ and $g$. Aside from a dilute gas of tightly-bound flux pairs, $V_{ij} = +1$ almost everywhere and most physical spins belong to the same ``domain'' once these flux pairs are ignored. As a result, we can construct a gauge-invariant computational two-point correlation function $\mathbb{G}(i,j)$ and a computational magnetization $\mathbb{M}$ which exhibit long-range order by first ``pairing'' Ising fluxes [see Fig.~\ref{fig:mwpm_numerics}(a)], thereby assigning each Ising spin to a well-defined domain.

To be precise, let $f_{ijk\ell} = V_{ij} V_{jk} V_{k\ell} V_{\ell i}$ denote the value of the flux through a plaquette $[ijk\ell]$, and let $f = \qty{f_{ijk\ell}}$ denote a particular configuration of fluxes. For each realization of $f$, we first compute a minimal-weight pairing $P[f] = \qty{P_{ij}[f]}$, which is a minimal-length set of paths through the dual lattice which connect fluxes\footnote{While $P$ always exists, it may be the case that $P$ is non-unique; in such a case, we simply define $\mathbb{G}(i,j)$ by implicitly averaging over all minimal-length pairings.}; it shall prove convenient to define $P_{ij} = -1$ ($P_{ij} = +1$) whenever the bond $\expval{ij}$ is included in (excluded from) the minimal-weight pairing. With this notation, $P$ is chosen to maximize the quantity $\sum_{\expval{ij}} P_{ij}$ subject to the constraint $P_{ij} P_{jk} P_{k\ell} P_{\ell i} = f_{ijk\ell}$ on each plaquette $[ijk\ell]$. Note that the product $VP = \qty{V_{ij} P_{ij}}$ of any domain-wall configuration and the minimal-weight pairing generated from its fluxes necessarily forms closed loops through the dual lattice. So long as these closed loops are homologically trivial, which we shall assume throughout this section (see Sec.~\ref{subsec:fs_bipartiteness} for a detailed discussion), they define a notion of domains with which one can construct correlation functions and an order parameter.

We now give four equivalent definitions of the computational two-point correlation function $\mathbb{G}(i,j)$, each of which provides a slightly different conceptual perspective of the observable: 


\textit{(1) Avoiding flux pairs:} for each domain wall configuration $V$, we read out a flux configuration $f$ and compute the corresponding minimal-weight pairing $P$. Using these, we construct a path $\gamma[f]$ from site $i$ to site $j$ which does not cross any of the links cut by the minimal-weight pairing (i.e., any link $\expval{ij}$ for which $P_{ij} = -1$). Such a path is always guaranteed to exist, since $P$ forms a collection of open strings in the dual lattice. We then define $\mathbb{G}(i,j)$ via
\begin{equation}
\label{eq:FS_G_comp}
    \mathbb{G}(i,j) = \expval{ \prod_{\expval{k\ell} \in \gamma[f]} V_{k \ell} },
\end{equation}
In this definition of $\mathbb{G}$, it is clear that the exponential decay arising in the low-temperature expansion of \eqref{eq:FS_string_corr} does not affect $\mathbb{G}$: each time a tightly-bound flux pair is inserted into the system, the string $\gamma[f]$ simply reorients to avoid this flux pair.

\textit{(2) Domain wall crossings:} if we now allow $\gamma[f]$ to be freely deformed from its original orientation, the value of $\mathbb{G}(i,j)$ will be maintained in each domain wall configuration $V$ if we introduce a minus sign each time $\gamma$ is deformed through a flux. Each such event changes the parity of the number of times $\gamma$ crosses a bond cut by the minimal-weight pairing. Therefore, an equivalent definition of $\mathbb{G}(i,j)$ is to allow for an arbitrary path $\gamma$, but to include an extra minus sign each time $\gamma$ crosses a bond in the minimal-weight pairing, resulting in the following:
\begin{equation}
\label{eq:FS_G_comp_2}
    \mathbb{G}(i,j) = \expval{ \prod_{\expval{k \ell} \in \gamma} V_{k \ell} P_{k \ell} } .
\end{equation}
Since $VP$ forms closed loops in the dual lattice, it is easy to see in this form that $\mathbb{G}(i,j)$ is manifestly path-independent, i.e., it depends only on $i$ and $j$ as the notation suggests. It is also clear from this definition that $VP$ should be regarded as defining the boundaries of the computationally constructed ferromagnetic domains.

\textit{(3) Zero-temperature random-bond Ising model:} so long as $VP$ forms homologically trivial closed loops, we can define spins $s_i$ within the domains obtained from $VP$ by writing $V_{ij} P_{ij} = s_i s_j$; this uniquely defines $s = \qty{s_i}$ up to an overall sign. Then, since the minimal-weight pairing $P$ is chosen to maximize $\sum_{\expval{ij}} P_{ij} = \sum_{\expval{ij}} s_i V_{ij} s_j$ for each domain wall configuration $V$, we can consider $s$ as the spin configuration arising from the zero-temperature ($\beta \to \infty$) limit of a random-bond Ising model (RBIM) with Hamiltonian
\begin{equation}
    H_{\text{RBIM}}[s,V] = -\beta \sum_{\expval{ij}} s_i V_{ij} s_j .
\end{equation}
Then, using \eqref{eq:FS_G_comp_2}, we immediately have our third equivalent formulation\footnote{In case the notation is confusing, recall that $\expval{\cdot}$ outside of the RBIM partition sum refers to an expectation value in the FS model; just as in Eqs.~\eqref{eq:FS_G_comp} and \eqref{eq:FS_G_comp_2}, Eq.~\eqref{eq:FS_G_comp_3} is to be regarded as a nonlocal computational observable of $H_{\FS}$.} of $\mathbb{G}(i,j)$:
\begin{equation}
\label{eq:FS_G_comp_3}
    \mathbb{G}(i,j) = \expval{ \lim_{\beta \to \infty} \frac{1}{Z_{\text{RBIM}}} \sum_{s} s_i s_j e^{-H_{\text{RBIM}}[s,V]} } ,
\end{equation}
where $Z_{\RBIM} = \sum_s e^{-H_{\RBIM}[s,V]}$. In words, we construct the computational two-point function $\mathbb{G}(i,j)$ by reinterpreting the domain wall variables $V_{ij}$ as quenched bond disorder in a RBIM, and computing the zero-temperature spin-spin correlation function in the resulting RBIM\footnote{Given this formulation of $\mathbb{G}$, expert readers might worry that $\mathbb{G}(i,j)$ exhibits spin-glass behavior at large $J$, rather than ferromagnetic behavior, as in the zero-temperature limit of the standard $\pm J$ RBIM. Numerically, we indeed find that the $g=0$ limit of this observable does not possess long-range ferromagnetic order, but any $g > 0$ appears to be sufficient to stabilize a ferromagnetic phase; see Fig.~\ref{fig:mwpm_numerics}(a).}. This formulation of $\mathbb{G}(i,j)$ will provide a useful point of comparison to the `intrinsic' computational observable to be defined in Sec.~\ref{subsec:fs_bipartiteness}.

\textit{(4) Minimal gauge:} since each of the previous definitions of $\mathbb{G}(i,j)$ is manifestly gauge-invariant, they are not explicitly defined in terms of the original spins $\sigma_i$. However, given the form of $\mathbb{G}$ in Eq.~\eqref{eq:FS_G_comp_3}, one might naturally ask whether the spins $s_i$ in the previous definition are related to the physical spins $\sigma_i$ in any meaningful way. It is easy to see that the spins $s_i$ exactly agree with $\sigma_i$ (up to a possible overall sign) when the latter are expressed in the \textit{minimal gauge}; i.e., we perform a gauge transformation so as to minimize the number of bonds on which $U_{ij} = -1$. Indeed, it is easy to see that $U_{ij}$ exactly agrees with $P_{ij}$ in this gauge, and so Eq.~\eqref{eq:FS_G_comp_2} immediately reduces to $\expval{\sigma_i \sigma_j}$ evaluated in the minimal gauge. While Ref.~\cite{fradkinPhaseDiagramsLattice1979} originally noted that spin-spin correlation functions in the minimal gauge of the FS model can exhibit long-range order for large $J$, the physical interpretation of this observation has remained unclear. Our minimal-weight pairing approach, and the three previous equivalent formulations of $\mathbb{G}(i,j)$, provide a gauge-invariant formulation of this observable.

Using any of the four preceding definitions of the computational correlation function $\mathbb{G}(i,j)$, we can now define a gauge-invariant computational magnetization $\mathbb{M}$, which is unique up to an overall sign. Fixing an initial site $i_0$ to define as spin-up, we define the sign of each other spin $\sigma_j$ relative to that of $\sigma_{i_0}$ by the number of times a path from $i_0$ to $j$ switches domains. Explicitly,
\begin{equation}
\label{eq:FS_M_comp}
    \mathbb{M} = \frac{1}{N} \sum_{j} \mathbb{G}(i_0, j) .
\end{equation}
Since $\mathbb{G}$ exhibits long-range order for large $J$, $\mathbb{M}$ is nonzero in this regime.

We now explicitly demonstrate our minimal-weight pairing protocol numerically. A detailed description of our numerical protocol is provided in Appendix~\ref{app_numerics}. In short, we first sample each domain wall configuration $V$ from the Hamiltonian~\eqref{eq:H_FS_2} via ordinary Metropolis Monte Carlo. Then, given each $V$, we compute a minimal-weight pairing of the fluxes $P$ using a sparsified version of the blossom algorithm \cite{edmondsPathsTreesFlowers1965,barahonaMorphologyGroundStates1982}, developed and implemented by Ref.~\cite{higgott2023sparse}. $V$ and $P$ together can be used to compute the computational two-point correlation functions $\mathbb{G}(i,j)$ and the computational magnetization $\mathbb{M}$ via Eqs.~\eqref{eq:FS_G_comp_2} and \eqref{eq:FS_M_comp} respectively. As we shall elaborate in Sec.~\ref{subsec:fs_bipartiteness}, this procedure yields well-defined computational observables only when the product $VP$ is homologically trivial; when topologically nontrivial boundary conditions are employed, we can check this by computing $\prod_{\expval{ij} \in \gamma} V_{ij} P_{ij} = \pm 1$ around each non-contractible cycle of the system. If this product is $-1$ for any of these cycles, then $\mathbb{G}(i,j)$ and $\mathbb{M}$ are simply ill-defined, and we set them to zero in this realization.

Figure~\ref{fig:mwpm_numerics}(b,c) showcases the most pertinent numerical results obtained from our minimal-weight pairing protocol. We focus here on $L \times L$ systems with periodic boundaries in both directions, and fix $g = 1$ in $H_{\FS}$ for simplicity. The results for other nonzero values of $g$ are qualitatively similar and exhibit quantitatively consistent critical exponents, as shown in Appendix \ref{app:more_numerics}.

Figure~\ref{fig:mwpm_numerics}(b) shows the computational magnetization $\mathbb{M}$ in the FS model, as a function of $J$ and for fixed $g=1$, for several system sizes. As anticipated, $\mathbb{M}$ serves as an order parameter which distinguishes between two computational phases. For large $J$ a computational ``ferromagnetic'' phase is realized, where $\mathbb{M}$ is finite; this indicates that the majority of spins belong to the same macroscopic cluster, as defined by minimal-weight pairing of Ising fluxes. In contrast, for small $J$ a computational ``paramagnetic'' phase is realized, where $\mathbb{M} = 0$ in the thermodynamic limit, indicating that the spins retain only short-range order whenever they are sorted into well-defined domains by minimal-weight pairing. These two regimes are separated by a non-thermodynamic continuous computational phase transition at a critical coupling $J_{c,\MW}(g=1) \simeq 0.6298$, as demonstrated by the excellent finite-size scaling collapse.

Our finite-size scaling analysis roughly estimates the critical exponents $\nu_{\MW} \simeq 1.4015$ and $\beta_{\MW} \simeq 0.1096$. It is useful to compare these critical exponents to several related models: namely, the clean $2d$ Ising model, the $\pm J$ RBIM at zero temperature, and bond percolation. Recall from Eq.~\eqref{eq:FS_G_comp_3} that the $g = 0$ limit of the computational magnetization $\mathbb{M}$ is exactly the magnetization in a standard $\pm J$ RBIM at zero temperature, with a given bond chosen to be negative with probability $p = [e^{2J} + 1]^{-1}$. Conversely, from either of Eqs.~\eqref{eq:FS_G_comp} or \eqref{eq:FS_G_comp_2}, the $g \to \infty$ limit  corresponds to the magnetization of a clean $2d$ Ising model at inverse temperature $J$. In the former case, a transition between paramagnetic and spin-glass phases is known to occur at $p_c \approx 0.103$ (corresponding to $J_c \approx 1.082)$, with a correlation length exponent of approximately $\nu \approx 1.46$ \cite{wangConfinementHiggsTransitionDisordered2003}. Our numerics suggest that $g > 0$ immediately eliminates the spin-glass phase in favor of a ferromagnetic phase; as might be expected, the critical coupling $J_{c,\MW}$ at $g = 1$ sits between the critical couplings of the standard RBIM at $g = 0$ and the clean Ising model at $g \to \infty$. While our estimated value of $\nu_{\MW}$ is consistent with those found numerically for the zero-temperature RBIM, it is also reasonably close to the bond percolation value $\nu = 4/3$ \cite{staufferIntroductionPercolationTheory2018}. Our estimate of the order parameter exponent $\beta_{\MW}$ is also relatively close to both the clean Ising exponent $\beta = 1/8$ and the bond percolation value $\beta = 5/36$. More detailed numerics are necessary to precisely establish the critical exponents and universality class of the computational transition in our model, which is beyond the scope of this work.

Finally, Fig.~\ref{fig:mwpm_numerics}(c) depicts the frequency of minimal-weight pairing failures, i.e., the frequency with which the minimal-weight pairing of fluxes resulted in a pairing $P$ which was non-homologous to the domain wall configuration $V$ (see Sec.~\ref{subsec:fs_bipartiteness}). Numerically, we find that this failure rate exhibits a transition at the same critical coupling $J_{c,\MW}$ as the computational magnetization, with an excellent finite-size scaling collapse using the same critical exponent $\nu_{\MW}$. For $J > J_{c,\MW}$, minimal-weight succeeds at pairing fluxes with unit probability in the thermodynamic limit; for $J < J_{c,\MW}$, minimal-weight fails with $75\%$ probability, due to the four inequivalent homology classes with periodic boundaries in both directions. In principle, it is possible to imagine that an intermediate \textit{third} phase could have existed between the computational ferromagnetic and paramagnetic phases, where minimal-weight pairing succeeds in pairing fluxes in the correct homology classes, but fails to establish a nonzero order parameter. Our numerics suggest that such a phase does not occur. Heuristically, the success of minimal-weight pairing of fluxes arises due to the nonzero domain wall tension, and the vanishing of the computational magnetization and the nonzero probability of minimal-weight pairing failures arise simultaneously as the domain wall tension vanishes. We demonstrate in Appendix~\ref{app:more_numerics} that using open boundary conditions, which exhibits no pairing failures, does not appear numerically to modify the critical point or the observed critical exponents.

\subsection{Bipartiteness Transition in the Fradkin-Shenker Model}
\label{subsec:fs_bipartiteness}

In defining $\mathbb{G}$ and $\mathbb{M}$ in the previous section, we have thus far been cavalier about boundary conditions. In an infinite system, or in a finite system with open boundary conditions, minimal-weight pairing always succeeds in constructing well-defined observables. However, close inspection shows that each of the four preceding definitions of $\mathbb{G}(i,j)$ are only well-defined when the minimal-weight flux pairing $P$ is \textit{homologous} to the domain wall configuration $V$; in the present context, $P$ and $V$ are homologous if and only if $P_{ij} = s_i V_{ij} s_j$ for a set of numbers $s_i = \pm 1$, indicating that $V$ can be sequentially deformed into $P$ by a series of spin-flips. When $V$ and $P$ are homologous, the product $VP$ forms homologically trivial closed paths through the dual lattice, which is a necessary condition for the products in Eqs.~\eqref{eq:FS_G_comp} and \eqref{eq:FS_G_comp_2} to be path-independent. In the presence of open boundary conditions, any two paths through the dual lattice with the same endpoints are homologous, and $\mathbb{G}(i,j)$ is always well-defined. This is no longer the case when topologically nontrivial boundary conditions, such as periodic or cylindrical boundary conditions, are employed: if $P$ and $V$ are non-homologous, then it is possible to complete a closed circuit through the lattice and switch domains an odd number of times. Therefore, if our minimal-weight pairing prescription chooses a pairing $P$ which is non-homologous to the domain wall configuration $V$, then the spins are not sorted into well-defined domains, and we say that our flux-pairing algorithm has failed. Since $\mathbb{G}(i,j)$ is ill-defined in this case, we simply set $\mathbb{G}(i,j) = 0$ in each domain wall configuration where $P$ is non-homologous to $V$.

In principle, there is likely to exist another protocol which can consistently pair fluxes in the correct homology class in a larger parameter regime than our previous minimal-weight algorithm; see Sec.~\ref{subsec:optimal} for one possible example of such an algorithm. However, as we will show in this section, for sufficiently small $J$ and $g$ \textit{no} algorithm can pair fluxes in the correct homology class with unit probability. Specifically, we show that there is a phase transition in the relative \textit{conditional} probabilities of each possible homology class, conditioned on the locations of the fluxes. For large $J$ and $g$, a given flux configuration $f$ is realized in a \textit{unique} homology class with unit probability in the thermodynamic limit, while for small $J$ and $g$ the probabilities of realizing the same flux configuration in different homology classes are comparable. In the latter phase, \textit{any} algorithm which attempts to establish a computational order parameter will necessarily fail with nonzero probability. The phase transition between these two regimes therefore provides an intrinsic, algorithm-independent computational distinction between the computational ferromagnetic and paramagnetic phases of the FS model. In analogy to the equivalent computational transition in AF melting, we shall refer to this transition as a ``bipartiteness'' transition, which separates a ``bipartiteable'' phase at large $J$ and $g$ and a ``non-bipartiteable'' phase at small $J$ and $g$.

To start, let us imagine that a particular domain wall configuration $V$ is sampled with probability $\mathcal{P}[V] \propto e^{-\beta H_{\FS}[V]}$ given by the usual Boltzmann weight. The domain wall configuration determines a particular flux configuration $f$, as well as a homology class $\hclass{V} = \qty{sVs}$, i.e., the equivalence class of all domain wall configurations $sVs = \qty{s_i V_{ij} s_j}$ homologous to $V$. With periodic boundary conditions in both spatial directions, there are four distinct homology classes corresponding to each flux configuration. Given a representative $V$ of the homology class $\hclass{V}$, we can obtain a representative $\bar{V}$ of a different homology class $\hclass{\bar{V}} \neq \hclass{V}$ by inserting a non-contractible domain wall around one of the cycles of the torus, i.e., by changing the sign $V_{ij} \to -V_{ij}$ along a non-contractible closed loop in the dual lattice.

We would like to determine if the homology class $\hclass{V}$ corresponding to a domain wall configuration $V$ can be determined from the flux configuration $f$ alone, with high probability in the thermodynamic limit. Towards this end, we shall compare the relative probabilities of two distinct homology classes with the same flux configuration. The probability $\mathbb{P}(\hclass{V})$ of sampling $\hclass{V}$ is obtained by simply summing $\mathcal{P}[V]$ over all domain wall configurations $V'_{ij} = s_i V_{ij} s_j$ in the homology class:
\begin{equation}
\label{eq:P(h)}
\mathbb{P}(\hclass{V}) \equiv \sum_s \mathcal{P}[sVs] \propto \sum_s \exp{J \sum_{\expval{ij}} s_i V_{ij} s_j} ,
\end{equation}
where we have neglected the flux fugacity term $e^{g VVVV}$ in the second expression, which depends only on the flux configuration. Quite naturally, the probability for the homology class $\hclass{V}$ is proportional to the partition function of a $\pm J$ RBIM with bond disorder determined by $V$. Each ``disorder realization'' is sampled with probability $\mathcal{P}[V]$. This should be compared to the computational two-point function $\mathbb{G}(i,j)$ defined via minimal-weight flux-pairing, specifically in the form \eqref{eq:FS_G_comp_3}; whereas the minimal-weight pairing algorithm exhibits the critical phenomena of a RBIM at \textit{zero} temperature, the computation of $\mathbb{P}(\hclass{V})$ exhibits the critical phenomena of a RBIM at a \textit{finite} temperature set by $J$.

For $g = 0$, each $V_{ij}$ is a statistically independent random variable, and $\mathbb{P}(\hclass{V})$ is precisely the partition function of the $\pm J$ RBIM on the Nishimori line \cite{nishimoriInternalEnergySpecific1981,nishimoriStatisticalPhysicsSpin2001a}. On the other hand, $g > 0$ introduces local correlations within the disorder which reduces the relative probability of bond configurations with Ising fluxes by $e^{-2g}$ per flux. The limit $g \to \infty$ results in a partition function which is gauge equivalent to a clean Ising model. Since both of these limits exhibit a continuous transition between ferromagnetic and paramagnetic phases tuned by $J$, it is natural to similarly expect a continuous phase transition in the ``partition functions'' \eqref{eq:P(h)}. We now construct several useful observables from these probabilities which can distinguish between the two phases, and which are operationally meaningful as computational observables of the original FS model.

First, to compare the relative probabilities of two distinct homology classes $h = \hclass{V}$ and $\bar{h} = \hclass{\bar{V}}$ which differ by a non-contractible domain wall, a natural observable is the relative entropy between the two probability distributions:
\begin{equation}
\label{eq:RBIM_F}
	\begin{split}
	    \mathbb{F} &= \sum_h \mathbb{P}(h) \qty[ \log \mathbb{P}(h) - \log \mathbb{P}(\bar{h}) ] \\
     &=\sum_V \mathcal{P}[V] \qty[ \log \mathbb{P}(\hclass{V}) - \log \mathbb{P}(\hclass{\bar{V}}) ] ,
	\end{split}
\end{equation}
where the latter expression follows from the former by noting that $\mathbb{P}(\hclass{V})$ is independent of the choice of representative $V$ in the homology class. From the second line, we see that $\mathbb{F}$ can be understood as the average free energy cost of inserting a non-contractible domain wall into a RBIM. In the original FS model, it is the \textit{conditional} free energy cost of inserting a non-contractible domain wall, conditioned on the locations of the fluxes. In the ferromagnetic phase of the RBIM, the insertion of a domain wall will typically cost an $\mathcal{O}(L)$ energy, and thus $\mathbb{F}$ will diverge with system size. This indicates that $\mathbb{P}(\hclass{V}) / \mathbb{P}(\hclass{\bar{V}})$ will approach either zero or infinity for typical $V$, suggesting that only one of the two homology classes has a finite probability in the thermodynamic limit. On the other hand, in the paramagnetic phase of the RBIM, the insertion of a domain wall will cost an $\mathcal{O}(1)$ energy, indicating that $\mathbb{P}(\hclass{V})$ and $\mathbb{P}(\hclass{\bar{V}})$ are comparably large for typical choices of $V$. 

Next, it is interesting to consider the free energy cost of inserting an additional unpaired flux into the system at plaquette $p$. While this cannot be done for a finite system with periodic boundary conditions in both directions, it is possible under \textit{cylindrical} boundary conditions, for example, with periodic boundaries in the $x$ direction and open boundaries in the $y$ direction. Then, an additional flux can be inserted into plaquette $p$ by changing the sign $V_{ij} \to -V_{ij}$  along a vertically oriented string in the dual lattice with endpoints at $p$ and the bottom boundary. We can obtain another useful computational observable by comparing the resulting domain wall configuration $V(\mu_p)$ to the original configuration $V$ as follows:
\begin{equation}
\label{eq:RBIM_mu}
    \expval{\mu_p} = \sum_V \mathcal{P}[V] \frac{\mathbb{P}(\hclass{V(\mu_p)})}{\mathbb{P}(\hclass{V})} .
\end{equation}
In the effective RBIM, this quantity is the usual definition of the Ising ``disorder parameter'' \cite{kadanoffDeterminationOperatorAlgebra1971a,fradkinDisorderOperatorsTheir2017}. As such, we expect that $\expval{\mu_p}$ will vanish in the ferromagnetic phase of the RBIM in the thermodynamic limit, but will be nonzero in the paramagnetic phase of the RBIM.

\begin{figure*}[t]
\includegraphics[width=\textwidth]{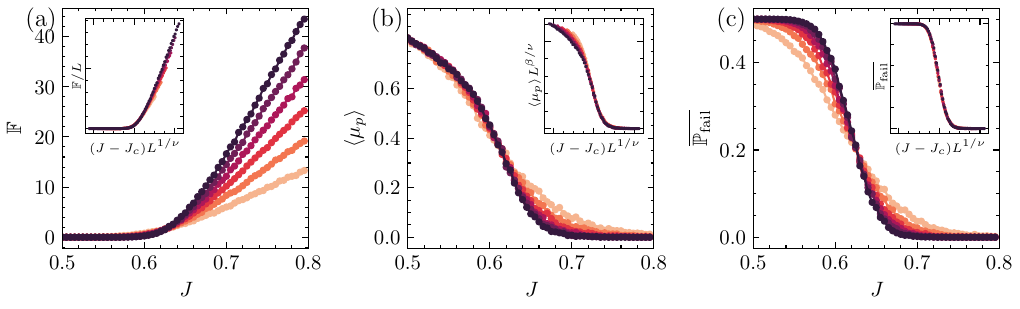}
\caption{Three ``intrinsic'' computational observables in the Fradkin-Shenker model [Eq.~\eqref{eq:H_FS}] at $g = 1$, computed using Monte-Carlo simulations and Gaussian fermion techniques, for square $L \times L$ lattices of various system sizes $L$ [see the legend in Fig.~\ref{fig:mwpm_numerics}(b)] with \textit{cylindrical} boundary conditions. (a) Conditional free energy cost of inserting a non-contractible domain wall into the system [Eq.~\eqref{eq:RBIM_F}], which exhibits $\mathcal{O}(L)$ scaling in the ``bipartiteable'' phase $J > J_{c,\opt}$ and $\mathcal{O}(1)$ scaling in the ``non-bipartiteable'' phase $J < J_{c,\opt}$. Inset: finite-size scaling collapse using the estimated critical point $J_{c,\opt} \simeq 0.6217$ and $\nu_{\opt} \simeq 1.6677$. (b) Ising ``disorder parameter'' $\expval{\mu_p}$, which represents the exponentiated free energy cost of inserting an additional unpaired flux into a plaquette $p$ in the center of the system. In the thermodynamic limit $\expval{\mu_p}$ vanishes in the bipartiteable phase and is finite in the non-bipartiteable phase. Inset: finite-size scaling collapse using the same estimated critical point $J_{c,\opt}$ and $\nu_{\opt}$ as in (a), as well as the additional critical exponent $\beta_{\opt} \simeq 0.1242$. (c) Probability $\overline{\mathbb{P}_{\text{fail}}}$ of failing to construct the computational observables $\mathbb{G}(i,j)$ and $\mathbb{M}$ by choosing a pairing in the most likely homology class [Eq.~\eqref{eq:RBIM_Pfail}]. In the thermodynamic limit, $\overline{\mathbb{P}_{\text{fail}}}$ vanishes in the bipartitable phase and is 0.5 in the non-bipartiteable phase with cylindrical boundary conditions.}
\label{fig:optimal_numerics} 
\end{figure*}

Finally, in order to construct computational observables such as $\mathbb{G}(i,j)$ and $\mathbb{M}$ in Eqs.~\eqref{eq:FS_G_comp} and \eqref{eq:FS_M_comp} respectively, we must choose one homology class $h$ in which to pair for each flux configuration $f$. If the various probabilities $\mathbb{P}(h)$ of these homology classes can be computed, then the ``optimal'' method for obtaining well-defined computational observables with the highest probability is achieved by simply selecting a pairing within the most likely homology class consistent with $f$. This prescription fails to produce a valid bipartitioning with probability
\begin{equation}
\label{eq:RBIM_Pfail}
    \mathbb{P}_{\text{fail}}(f) = 1 - \frac{\text{max}_{h | f}[\mathbb{P}(h)]}{\sum_{h | f} \mathbb{P}(h)} .
\end{equation}
where the notation $h | f$ in the numerator (denominator) denotes a maximum (summation) over all homology classes $h$ consistent with the flux configuration $f$. We will also denote the \textit{average} failure probability as $\overline{\mathbb{P}_{\text{fail}}}$. In the ferromagnetic phase of the RBIM, $\overline{\mathbb{P}_{\text{fail}}}$ is exponentially small in $L$, while in the paramagnetic phase $\overline{\mathbb{P}_{\text{fail}}}$ saturates at $3/4$ for a system defined on the torus. On a cylinder, $\overline{\mathbb{P}_{\text{fail}}}$ instead saturates at $1/2$ in the paramagnetic phase.

In summary, we have found that there is a phase transition in the relative probabilities for the homology classes of domain walls for a given flux configuration. For large $J$, only one of the four possible homology classes has nonzero probability in the thermodynamic limit, while for small $J$ each of the four classes can occur with relatively similar probabilities. In the large $J$ phase, given a flux configuration $f$, we can determine the ``correct'' homology class by numerically computing the partition functions $\mathbb{P}(h)$ for each homology class $h$ consistent with $f$ and choosing the most probable one.

We now demonstrate the proposed computational transition numerically. As in Sec.~\ref{subsec:fs_mwpm}, we start by sampling a domain wall configuration $V$ via Metropolis Monte Carlo. Given $V$, we then compute the RBIM partition function $\mathbb{P}(\hclass{V})$ \textit{exactly} using Gaussian fermion techniques \cite{bravyiLagrangianRepresentationFermionic2004,merzTwodimensionalRandombondIsing2002,bravyiEfficientAlgorithmsMaximum2014}, as reviewed in Appendix~\ref{app_numerics}. It is technically convenient in this section to work with \textit{cylindrical} boundary conditions, i.e., periodic boundaries in the $x$ direction and open boundaries in the $y$ direction; not only does this simplify the numerical computation of $\mathbb{P}(\hclass{V})$, but it also allows for the disorder parameter $\expval{\mu_p}$ to be defined. Figure \ref{fig:optimal_numerics} depicts the most pertinent numerical results for probing the bipartiteness transition in the FS model. As in Sec.~\ref{subsec:fs_bipartiteness}, we focus on $L \times L$ systems with $g = 1$, but now employ cylindrical boundary conditions; as a result, there are only two distinct homology classes in this geometry. 

Figure \ref{fig:optimal_numerics}(a) depicts the conditional free energy $\mathbb{F}$ of inserting a non-contractible domain wall, as defined in Eq.~\eqref{eq:RBIM_F}. As expected, $\mathbb{F}$ scales linearly with $L$ for large $J$. This indicates that, in the thermodynamic limit, only one of the two homology classes has a finite probability for a given flux configuration; the probability of the rarer homology class vanishes exponentially with $L$. For small $J$, $\mathbb{F}$ is of order unity, indicating that the two homology classes for a given flux configuration are comparably probable. The two regimes are separated by a continuous computational phase transition at a critical coupling strength $J_{c,\opt}(g=1) \simeq 0.6217$. This value is slightly lower than the critical coupling strength $J_{c,\MW}$ obtained from minimal-weight pairing in Sec.~\ref{subsec:fs_mwpm}, consistent with the interpretation that minimal-weight pairing is sometimes suboptimal in choosing the correct homology class in which to pair. Note, however, that $J_{c,\MW}$ and $J_{c,\opt}$ are very close, and minimal-weight pairing is expected to perform excellently for all practical purposes. 

It is also interesting to note that the estimated correlation length exponent $\nu_{\opt} \simeq 1.6677$ is appreciably larger than the one obtained in the minimal-weight pairing transition, although more detailed numerics are required to definitively claim that these exponents are different from one another. The large value of $\nu_{\opt}$ is somewhat unexpected, since the clean Ising transition with $\nu = 1$ is recovered in the limit $g \to \infty$, while the Nishimori transition with $\nu \approx 1.48$ \cite{merzTwodimensionalRandombondIsing2002,piccoStrongDisorderFixed2006} is recovered in the limit of $g \to 0$. Naively, one expects that $\nu_{\opt}$ should either lie somewhere between $1$ and $1.48$, or flow all the way to the clean Ising value of 1 for nonzero $g$. Larger-scale numerics are required to confidently establish the observed value of $\nu_{\opt}$, which is again beyond the scope of this work.

Figure \ref{fig:optimal_numerics}(b) shows the ``disorder parameter'' defined in Eq.~\eqref{eq:RBIM_mu}, for a plaquette $p$ in the center of the system. We see that $\expval{\mu_p}$ vanishes in the thermodynamic limit in the bipartiteable phase and is finite in the non-bipartiteable phase, as expected. Using $\expval{\mu_p}$, we can extract an additional critical exponent $\beta_{\opt} \simeq 0.1242$. Although this estimated exponent is very close to the clean Ising value $\beta = 1/8$, the large value of $\nu_{\opt}$ suggests that the transition may lie in a different universality class. 

Finally, Fig.~\ref{fig:optimal_numerics}(c) depicts the rate of failures $\overline{\mathbb{P}_{\text{fail}}}$, in which the sampled domain wall configuration $V$ does not fall into its flux configuration's most likely homology class. As predicted, $\overline{\mathbb{P}_{\text{fail}}}$ vanishes in the thermodynamic limit for all $J > J_{c,\opt}$ and sharply transitions to $0.5$ for $J < J_{c,\opt}$. This once again establishes that fluxes can always be consistently paired to establish computational observables in the bipartiteable phase, while in the non-bipartiteable phase computational observables such as the computational magnetization $\mathbb{M}$ are ill-defined. The excellent finite-size scaling collapse indicates that the observed value of $\nu_{\opt}$ is consistent across all three observables presented here.

\subsection{Conjectured Optimal Pairing Algorithm}
\label{subsec:optimal}

The bipartiteness transition described in the previous section places a fundamental restriction on any algorithm which attempts to construct a computational order parameter in the FS model via flux-pairing in the presence of topologically nontrivial boundary conditions. In the non-bipartiteable phase, since the homology class of the domain wall cannot be inferred from its flux configuration with unit probability, no flux-pairing algorithm can reliably construct a well-defined order parameter. As a result, we expect that the computational magnetization $\mathbb{M}$ arising from any flux-pairing algorithm will vanish in the thermodynamic limit. 

A natural question is whether one can establish a similar ``intrinsic'' computational phase transition without appealing to the system's global boundary conditions, which presumably should not affect the presence or absence of magnetic order. Indeed, flux-pairing is always guaranteed to establish consistent computational observables in the presence of open boundary conditions, but we nevertheless expect that no flux-pairing algorithm can establish a nonzero computational magnetization for sufficiently small $J$. In this section we construct a flux-pairing algorithm which we conjecture to be ``optimal'', in the sense that it establishes a nonvanishing computational magnetization in the largest possible parameter regime. Although we do not attempt to rigorously prove the optimality of our algorithm, we present highly plausible physical arguments which suggest that no other flux-pairing algorithm can establish a nonzero computational magnetization when our proposed algorithm fails to do so. The phase transition in the proposed algorithm therefore provides an intrinsic phase boundary between computational ferromagnetic and paramagnetic phases, independent of global boundary conditions.

To state the algorithm, recall in Sec.~\ref{subsec:fs_mwpm} that we chose a pairing $P[f]$ from a given flux configuration $f$ by demanding that $P$ had minimal weight, i.e., that $\sum_{\expval{ij}} P_{ij}$ was maximized. Here we instead propose to \textit{randomly} select the pairing $P$ with the following probability:
\begin{equation}
    \mathcal{P}_{\opt}(P) = \frac{1}{Z_{\opt}[f]} e^{J \sum_{\expval{ij}} P_{ij}}, \ Z_{\opt}[f] = \sum_{P | f} e^{J \sum_{\expval{ij}} P_{ij}} ,
\end{equation}
where the latter sum is performed only over pairings $P$ consistent with the flux configuration $f$. To see why this random pairing algorithm is conjectured to be optimal, consider first the case of open boundary conditions. Then, since $P$ and $V$ are guaranteed to be homologous, we may write $P_{ij} = s_i V_{ij} s_j$ as in Sec.~\ref{subsec:fs_mwpm}. Averaging over realizations of $P$, the computational two-point function $\mathbb{G}_{\opt}(i,j)$ obtained from this algorithm can be written as
\begin{equation}
\label{eq:G_comp_opt}
    \begin{split}
        \mathbb{G}_{\opt}(i,j) &= \expval{ \frac{1}{Z_{\opt}[f]} \sum_{P | f} \left[ \prod_{\expval{k\ell} \in \gamma} V_{k\ell} P_{k\ell} \right] e^{J \sum_{\expval{ij}} P_{ij}} } \\
        &= \expval{ \frac{1}{2Z_{\opt}[f]} \sum_{s} s_i s_j e^{J \sum_{\expval{ij}} s_i V_{ij} s_j} } ,
    \end{split}
\end{equation}
where $\gamma$ in the first line is an arbitrary path from site $i$ to site $j$, and the factor of 2 in the denominator of the second line corrects the overcounting of pairings $P$ due to the two-to-one mapping from $s$ to $P$. The denominator can similarly be written as
\begin{equation}
    Z_{\opt}[f] = \frac{1}{2} \sum_{s} e^{J \sum_{\expval{ij}} s_i V_{ij} s_j} .
\end{equation}
Thus, $\mathbb{G}_{\opt}$ can be understood as a two-point correlation function in a RBIM, with bond disorder $V$ distributed according to the Boltzmann weight $e^{-H_{\FS}[V]}$. In fact, this is an \textit{identical} RBIM to the one derived for the bipartiteness transition in Sec.~\ref{subsec:fs_bipartiteness}, except for the assumption of open boundary conditions in the present section\footnote{More generally, when topologically nontrivial boundary conditions are employed, the results of Sec.~\ref{subsec:fs_bipartiteness} imply that the pairings $P$ which are non-homologous to $V$ arise with vanishing probability in the thermodynamic limit. Therefore, the change of variables from $P$ to $s$ in Eq.~\eqref{eq:G_comp_opt} is valid throughout the entire bipartiteable phase for \textit{any} boundary conditions.}. Therefore, as long as the global boundary conditions of this RBIM do not affect the location or universality of its phase transition, we expect that our proposed pairing algorithm will successfully construct a computational magnetization $\mathbb{M}_{\opt} = \frac{1}{N} \sum_j \mathbb{G}_{\opt}(i_0, j)$ which is nonzero throughout the entire bipartiteable phase and zero throughout the entire non-bipartiteable phase. So long as no other flux-pairing algorithm can construct a nonzero computational magnetization in the non-bipartiteable phase, $\mathbb{M}_{\opt}$ is nonzero in the largest possible parameter regime.

\section{Hard-Sphere Monte Carlo Numerics}
\label{sec:hardspheres}

In Secs. \ref{subsec:fs_mwpm}-\ref{subsec:optimal}, we studied computational phase transitions in the Fradkin-Shenker model as an \textit{effective} model for the Ising spins within the AF/PM tetratic phase. In this section, we illustrate how the ideas presented above are manifested in a concrete physical system. Consider a system of \textit{hard-sphere} colloids of uniform diameter, $\sigma$, confined to move between horizontal plates separated by a height $H$ \cite{Yair_prl_frustration_ising_colloids,Han_nature_frustration_ising_colloids,abutbulTopologicalOrderAntiferromagnetic2022}.  Hence, the colloids are free to move in the lateral ($xy$) directions, but have limited room to move in the vertical direction. The colloids are suspended in a fluid with matching density, such that the effects of gravity are canceled, and matching dielectric constant, such that there are no interactions between colloids other than the hard-core constraint preventing their overlap.  The phase diagram of the system can then be tuned by two dimensionless parameters, the normalized density $\rho_H=\frac{N\sigma^3}{AH}$ and plate separation $h=\frac{H}{\sigma}-1$.  Here, $N$ is the number of colloids and $A$ is the system area.

We restrict our attention to $h<1$, so that it is not possible to stack colloids on top of each other.  Then, the height of the colloids relative to the central plane, $z$, can be thought of as encoding an Ising spin degree of freedom. At nonzero temperature, entropic forces favor configurations in which colloids are well-separated from each other, thus increasing the spatial fluctuations available to them. Hence, nearby colloids prefer to lie near opposite confining plates, corresponding to an effective AF interaction between neighboring spins. 

This system was simulated numerically in Ref.~\onlinecite{abutbulTopologicalOrderAntiferromagnetic2022} using the Event Chain Monte Carlo method \cite{Bernard_ECMC,Bernard_two_step_ECMC} on systems containing up to $N=90,000$ colloids.  For $h=0.8$, the phase diagram consists of an antiferromagnetic solid phase for $\rho_H\gtrsim 0.84,$ a tetratic phase for $0.78\lesssim \rho_H\lesssim 0.84$, and a liquid phase for $\rho_H\lesssim 0.77$ \cite{abutbulTopologicalOrderAntiferromagnetic2022}.  The AF solid was identified from the observed power-law positional correlations, long-range orientational order, and power-law N\'eel correlations, corresponding to algebraically divergent Bragg peaks in the magnetic structure factor. The tetratic phase displayed exponentially-decaying positional correlations and power-law orientational correlations.  The liquid phase had exponentially decaying positional and orientational correlations.  For $0.775\lesssim \rho_H\lesssim 0.78$, a bimodal distribution in the histogram of orientational order parameter values was observed, as well as a proliferation of grain boundaries, both of which are possible indications of a tetratic/liquid coexistence region and a first-order transition between these phases. 

The tetratic phase was demonstrated to have free double dislocations despite the fundamental dislocations being bound \cite{abutbulTopologicalOrderAntiferromagnetic2022}.  This phase showed strong AF {\em correlations}, resulting in broadened magnetic Bragg peaks at the N\'eel vector of the putative magnetic order.  The broadening of the magnetic Bragg peaks reflects the short-range positional correlations in the tetratic phase, which prevent the N\'eel order parameter from achieving a nonzero expectation value irrespective of the arrangement of the spins.  

Here, our goal is to search for {\em long-range computational AF order} in the tetratic phase and, if present, to quantify it. For this, we define a staggered magnetization order parameter by \textit{algorithmically} dividing the colloids into two sublattices, as illustrated schematically in Fig.~\ref{fig:pairing_protocol}, and computing the difference in Ising spins between them, as in Eq.~(\ref{eq:stag_mag_2}).  However, unlike the lattice model considered in Secs.~\ref{subsec:fs_mwpm}-\ref{subsec:optimal}, the colloids are free to move in the continuum.  Therefore, the construction of the order parameter requires a number of additional steps, which we now describe.  

\begin{figure}[t]
	\centering
	\includegraphics[width=\columnwidth]{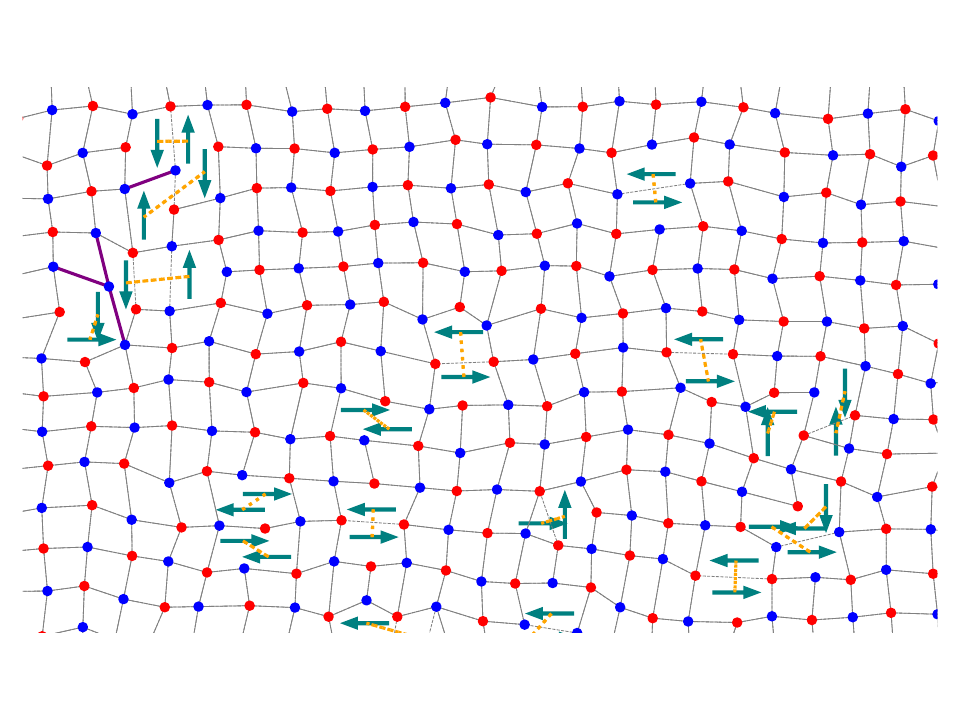}
	\caption{Hard-sphere configuration in a snapshot of the Monte-Carlo simulation in the tetratic phase ($\rho_H=0.81$ and $h=0.8$).  A small region of the total simulation (containing $N=90,000$ spheres) is shown. The spheres are colored blue/red if they are above/below the center plane (i.e., according to their ``spin'').  Green arrows are dislocations computed using DXA.  Gray lines connect spheres that satisfy our nearest-neighbor condition.  Orange dotted lines join pairs of nearby dislocations (pairs are determined through an algorithm that minimizes total distance between pairs).  Gray bonds crossing orange dotted lines are removed from the graph, making it bipartite. The AF order parameter is the difference of total spin between the two subgraphs of the resulting bipartite graph.}
	\label{fig:dislocation-pairing}
\end{figure}

The first step is to locate the dislocations in the system following the dislocation extraction analysis (DXA) in Ref.~\onlinecite{Stukowski_dxa}.
For this, we first rotate the system to align the average orientation of nearest-neighbor bonds with the $x$-axis. We then perform a Delauney triangulation of the sphere lateral positions and compare the links of the triangulation to those of a reference perfect square lattice, i.e., vectors of the form ${\bf v}_{n_1,n_2}=a \left(n_1 \hat{\vb{x}}+n_2\hat{\vb{y}}\right)$, where $n_1$ and $n_2$ are integers and $a$ is the lattice constant obtained from the peak in the static structure factor. We restrict $n_1,n_2\in \{-1,0,1\}$, i.e., to nearest-neighbor and next-nearest neighbor links of the reference perfect square lattice. Then, each link ${\bf w}$ of the Delauney triangulation is assigned the link on the reference lattice that minimizes the distance $|{\bf w}-{\bf v}_{n_1,n_2}|^2$. The Burgers vector on each triangle of the triangulation is then computed by building a Burgers circuit on the triangle using the links of the reference lattice.  Figure~\ref{fig:dislocation-pairing} shows the dislocation field on a snapshot of the Monte-Carlo simulation.

A criterion for spheres to be considered neighbors is defined based on a similar procedure.  Starting from the Delauney triangulation, we remove the links that were assigned next-nearest neighbor bonds of the reference square lattice.  Spheres connected by the remaining links are then considered nearest neighbors. This criterion differs from that used in \cite{abutbulTopologicalOrderAntiferromagnetic2022}; it has the advantage of guaranteeing a direct relationship between the location of the dislocations of the DXA and the Burgers circuits of the resulting graph.

The end result is a graph of points connected to their neighbors, which locally resembles a square lattice, except for the effect of phonons and dislocations, see Fig.~\ref{fig:dislocation-pairing}. In particular, the graph is not globally bipartite due to the presence of fundamental dislocations, which tend to appear in close-by pairs.  As described in Sec.~\ref{subsec:comp_overview}, the graph can be made bipartite by grouping fundamental dislocations in pairs, connecting the dislocations in each pair by a line, and removing all links that cross that line.  It then becomes impossible to make an odd Burgers circuit, since a circuit must enclose fundamental dislocations in pairs.  When open boundary conditions are used, this guarantees that the lattice is bipartite.  With periodic boundary conditions, on the other hand, bipartiteness may still fail: when the pairing paths and the physical domain walls form together a non-contractible cycle around the torus, as shown in Fig.~\ref{fig:pairing_protocol}(c), loops wrapping around the torus in the other direction can be odd.  Inside the AF tetratic regime, the probabilty of such odd loops is vanishing in the thermodynamic limit; on the other side of the computational bipartiteness transition, they have non-zero probability.  The AF order parameter can only be defined on one side of the computational transition.

Similar to the minimal-weight pairing algorithm used in the lattice gauge theory, we pair dislocations here using a \textit{minimal-distance} (MD) protocol, in which we minimize the sum of the Euclidean distances between paired dislocations.  For this, we use a minimum-weight matching algorithm on a graph with dislocations as vertices, and with edges whose weight is the distance between dislocations. 

\begin{figure}
     \centering
\subfloat{
\begin{overpic}
[width=.48\textwidth]{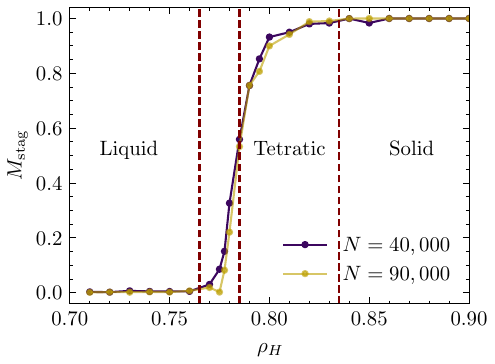}
\put(1, 63){\large (a)}
\put(41,52){L/T}
\end{overpic} 
\label{fig:AF_OPa}}\\
\subfloat{\begin{overpic}[width=.48\textwidth]{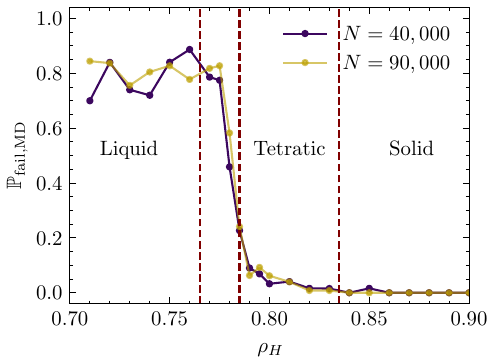}
\put(1, 63)
{\large (b)}
\put(41,64){L/T}
\end{overpic}
\label{fig:AF_OPb}}
\caption{(a) Antiferromagnetic order parameter in a system of hard spheres with plate separation $h=0.8$. L/T indicates a liquid/tetratic coexistence region. (b) Probability that minimal-distance dislocation pairing fails to yield a bipartite graph in systems with periodic boundary conditions.}
	\label{fig:AF_OP}
\end{figure}

Figure \ref{fig:AF_OP}(a) shows the order parameter $M_{\mathrm{stag}}$ as a function of colloid density for $N=90,000$ and $N=40,000$ particles.  As can be seen, in the solid phase, the AF order saturates close to its maximal value, $M_{\mathrm{stag}}\approx 1$.  Interestingly, as $\rho_H$ is reduced, the staggered magnetization shows no clear signature upon entering the tetratic phase. As $\rho_H$ is reduced further, the staggered magnetization is suppressed, until it approaches zero. 
 This suppression accelerates upon approaching the tetratic/liquid coexistence region, and it also becomes more strongly system-size dependent, indicating that in the thermodynamic limit the order parameter is likely to vanish in the coexistence region.

The simulations were performed with periodic boundary conditions, allowing us to test the bipartiteness of graphs following dislocation pairing.  Figure \ref{fig:AF_OP}(b) shows the bipartiteness failure probability, $\mathbb{P}_{\text{fail,MD}}$, for dislocation pairing based on the minimal distance protocol.  The failure probability remains small throughout the tetratic phase, and grows significantly upon entering the liquid/tetratic coexistence region. We estimate the computational bipartiteness transition transition to occur at a critical density $\rho_H^{\mathrm{c,MD}}\approx 0.785$. There is large uncertainty in this estimate, however, due to the large statistical fluctuations in the failure probability which overwhelm the system-size dependence within the tetratic phase.     We emphasize that the transition in question is specific to the minimal-distance algorithm -- we expect an intrinsic bipartiteness transition to occur at a lower critical density.

\section{Discussion}
\label{sec:discussion}

In this work, we have demonstrated two types of computational phase transitions which arise naturally in the context of two-dimensional classical AF melting. Specifically, in order to construct an order parameter which can sharply distinguish between the AF and PM tetratic regimes, one must first algorithmically bipartition the atoms into two sublattices by pairing dislocations. One particularly simple algorithm for this task is minimal-weight pairing, which minimizes the number of nearest-neighbor atoms which are grouped into the same sublattice; this algorithm can be efficiently implemented \cite{edmondsPathsTreesFlowers1965,barahonaMorphologyGroundStates1982}, and the resulting AF order parameter undergoes a sharp computational phase transition between the AF and PM tetratic regimes. More generally, employing topologically nontrivial boundary conditions reveals an \textit{intrinsic} computational phase transition in the bipartiteness of the lattice: for sufficiently strong AF coupling, bipartiteness can be restored to the atoms in an essentially unique way, while below a critical threshold \textit{no} algorithm can consistently bipartition the atoms when periodic boundary conditions are imposed.

It is interesting to note the close analogy between our dislocation-pairing protocol and quantum error correction protocols in topological codes \cite{dennisTopologicalQuantumMemory2002,bombinIntroductionTopologicalQuantum2013}, particularly in the simple case of bit-flip error channels on the toric code. In this setting, a set of bit-flip errors occurs independently on each qubit with probability $p$, taking the system out of its code space by introducing a number of $m$ anyons. The positions of these anyons are measured, and one attempts to infer the homology class of the underlying error, i.e., an equivalence class of errors which differ only by stabilizer operations. If this homology class can be determined with unit probability in the thermodynamic limit, then the errors can always be corrected without introducing a logical operation on the code space, and the error rate $p$ is said to be below the error threshold. In our model, we can consider domain walls as analogous to bit-flip errors and elementary dislocations as analogous to $m$ anyons. Working with periodic boundary conditions, if domain walls for a given configuration of dislocations belong to a unique homology class, then it is always possible to pair these dislocations and establish a bipartite structure on the lattice, and thereby define a staggered magnetization. On the other hand, if a given configuration of dislocations can correspond to two homologically inequivalent domain wall configurations with comparable probabilities in the thermodynamic limit, then a staggered magnetization cannot be consistently defined. 

One important difference between our model and the error correction problem in the toric code is that bit-flip errors in the latter are each sampled independently, while domain wall configurations in the former are sampled according to the full Boltzmann weight of the Hamiltonian \eqref{eq:ham_schematic}. As a result, the ``errors'' in our model exhibit local correlations, which can potentially modify the universality class of the threshold transition. Indeed, whereas the transition in the case of independent errors is described by the Nishimori point of the $\pm J$ RBIM \cite{dennisTopologicalQuantumMemory2002}, the critical exponents observed in Sec.~\ref{subsec:fs_bipartiteness} suggest that the bipartiteness transition in the present work may belong to a distinct universality class. Nevertheless, more detailed numerics are required to convincingly distinguish the exponents found here from those of similar models in the literature.

Another important difference between dislocation-pairing and anyon-matching is that homologically equivalent pairings in the latter problem are entirely indistinguishable, while two homologous dislocation-pairings can lead to observably different AF order parameters. This is not expected to make a significant difference in the thermodynamic limit, where any \textit{finite} deformation in the definition of domains cannot eliminate a nonzero staggered magnetization. However, this observation does imply that simply identifying the correct homology class of the domain walls does not by itself furnish an ``optimal decoding'' of the AF order. To promote our homology-identifying algorithm to a (generally non-optimal) ``decoder'', it must be supplemented by an algorithm which chooses a particular pairing of dislocations within the correct homology class. For example, one can introduce a modified minimal-weight pairing algorithm where the minimal-weight pairing within a particular homology class is employed; or, the conjectured optimal algorithm of Sec.~\ref{subsec:optimal} can be modified to sample pairings only within the correct homology class.

Aside from connections to quantum error correction, it is interesting to ask whether the observed computational transitions are related to other known examples of non-thermodynamic phase transitions in statistical physics. For example, in a two-dimensional Ising model subjected to a symmetry-breaking field $h$, clusters of spins are known to undergo a non-thermodynamic percolation transition along a line in the $(J,h)$ plane known as the Kert\'{e}sz line \cite{kerteszExistenceWeakSingularities1989b,blanchardKertEszLine2008}; as might be expected, the FS model \eqref{eq:H_FS} similarly exhibits a Kert\'{e}sz line despite having no thermodynamic phase transition \cite{nussinovDerivationFradkinShenkerResult2005}. It is a priori unclear where this Kert\'{e}sz line sits in relation to the computational phase transitions discussed in this work, but given the physical connection to percolating clusters of spins, it would be unsurprising if these transitions were quite close in parameter space.

Moving away from computational phase transitions, there remains many intriguing questions about the thermodynamic phase diagram of the AF melting problem. One immediate question is regarding the interplay between AF order and \textit{disclinations}, topological defects in the orientational order, which become energetically allowed within the tetratic phase. In square lattices, there turn out to be two topologically inequivalent types of disclinations \cite{gopalakrishnanDisclinationClassesFractional2013}. One of these grossly violates the bipartiteness of the lattice similar to elementary dislocations, while the other maintains the lattice's bipartiteness. It is interesting to consider whether the proliferation of the latter disclinations can result in an AF liquid \cite{timmLiquidAntiferromagnetsTwo2002}, which would presumably exhibit a non-thermodynamic computational phase transition to a PM liquid as elementary dislocations and bipartiteness-violating disclinations unbind.

Another important thermodynamic question is regarding the absence of a thermodynamic transition separating the AF and PM tetratic phases. In Sec.~\ref{sec:noPT} we explained that the unbinding of elementary dislocations eliminated the sharp phase transition separating AF and PM order within the tetratic phase. It is natural to ask whether there can exist another model of AF melting which does contain a sharp thermodynamic phase transition between the AF and PM tetratic phases. Indeed, Ref.~\cite{shamai2018molten} demonstrated exactly such a transition between AF and PM hexatic phases of a buckled Coulomb crystal \cite{podolskyBucklingTransitionsClock2016}, where dislocations bind to $\pm \frac{1}{3}$ fractional vortices in an effective six-state clock model. We revisit this AF/PM hexatic transition in Appendix~\ref{app_buckled_crystal} using the perspective of Sec.~\ref{sec:noPT}, finding (consistent with the results of Ref.~\cite{shamai2018molten}) that dislocations eliminate the XY phase of the clock model but maintain a sharp Ising-like transition between the AF and PM hexatic phases. 

Due to the high complexity of microscopic models of classical melting, it is helpful to identify a simple lattice model which exhibits similar phenomenology to the AF melting problem discussed in this work. The simplest such model is the modified XY model introduced by Korshunov, Lee, and Grinstein \cite{leeStringsTwodimensionalClassical1985a,korshunovPhaseDiagramModified1986}. In addition to the usual integer vortex topological excitations, the XY phase of this model exhibits \textit{half-integer} vortices which are linearly confined at the endpoints of string-like Ising domain walls. These domain walls proliferate in a $2d$ Ising transition to a ``pair-superfluid'' phase, resulting in only a logarithmic confinement of half-vortices. Quite remarkably, Ref.~\cite{shiBosonPairingUnusual2011} identified a parameter regime where the transition from the XY phase to the disordered phase proceeds through an Ising-like transition, rather than via a Kosterlitz-Thouless transition. By identifying half-vortices (integer vortices) with elementary dislocations (double dislocations), it is natural to postulate that a similar Ising transition from AF solid to PM tetratic may arise in the present work's model as well. Furthermore, the modified XY model is known to exhibit a non-thermodynamic ``deconfinement transition'' within its disordered phase \cite{sernaDeconfinementTransitionsGeneralised2017}, roughly corresponding to the deconfinement of half-vortices. By analogy to the dislocation-pairing problem of the present work, one naturally suspects that there might also exist a computational phase transition in an algorithm which attempts to pair half-vortices.

An immediate generalization of the present work is to consider the \textit{quantum} antiferromagnetic tetratic at zero temperature. In this context, the computational observables defined in this work have an intimate relation to measurement-induced phenomena such as in Refs.~\cite{Garratt2023Measurements,weinsteinNonlocalityEntanglementMeasured2023a,lee2022decoding,luMixedStateLongRangeOrder2023a}. Specifically, while true long-range AF order is expected to be absent in the tetratic phase as in the classical case, we expect that AF order can be ``decoded'' from the positions of dislocations. First, the positions of the atoms are measured and the elementary dislocation ``syndromes'' are identified. Then, following a pairing procedure analogous to the ones described in this work, the atoms are sorted into $A$ and $B$ sublattices and spins in sublattice $B$ are flipped along the quantization axis. If the resulting quantum state exhibits long-range \textit{ferromagnetic} order, then we have successfully decoded the AF order. Just as in the classical case, this error correction protocol promotes the smooth crossover between AF and PM tetratic regimes to a sharp phase transition. It is interesting to ask more generally how error correction, or local measurements combined with feedback and nonlocal classical communication, can lead to novel quantum phases or phase transitions which do not arise naturally in the ground state of a local Hamiltonian.

\acknowledgments
\textit{Acknowledgments.--} We gratefully acknowledge Daniel Abutbul, Sajant Anand, Stefan Divic, Ruihua Fan, Sam Garratt, Sarang Gopalakrishnan, Jaewon Kim, Zohar Nussinov, Akshat Pandey, and Andrew Potter for discussions and insightful comments.  This work was supported by in part by a Simons Investigator Award (E.A.), NSF QLCI program through Grant No. OMA-2016245 (E.A. and Z.W.), and the Israel Science Foundation under grant No.~2541/22 (D.P. and J.A.A.).

\onecolumngrid
\appendix

\section{Numerical Details}
\label{app_numerics}
In this Appendix, we provide additional details on the numerical calculations employed in Secs.~\ref{subsec:fs_mwpm} and \ref{subsec:fs_bipartiteness}. Specifically, we first briefly mention our Monte Carlo algorithm for sampling domain wall configurations $V$ and how minimal-weight pairing of fluxes is performed, and then explain in detail the method by which various RBIM partition functions are computed in Sec.~\ref{subsec:fs_bipartiteness}.

\subsection{Monte Carlo Method and Minimal-Weight Pairing}
In order to sample domain wall configurations $V = \qty{V_{ij}}$ from the probability distribution $\mathbb{P}[V] \propto e^{- H_{\FS}[V]}$ [see Eq.~\eqref{eq:H_FS_2}], we employ a standard Metropolis algorithm \cite{landauGuideMonteCarlo2014}. During each Monte Carlo step, a domain wall configuration $V$ is evolved to a new configuration $V'$ via two distinct Metropolis updates: 
\begin{enumerate}
    \item First, we choose a bond $V_{ij}$ and flip its sign with probability $e^{-\Delta}$;
    \item Second, we choose a site $i$ at random and simultaneously flip the sign of all four bonds $V_{ij}$ containing $i$ with probability $e^{-\Delta}$.
\end{enumerate}
In both cases, $\Delta \equiv H_{\FS}[V'] - H_{\FS}[V]$ is the change in energy following the update. If $\Delta < 0$ (i.e., the update lowers the system energy), the update is accepted with unit probability. The second update step is especially important when $g$ is large: since the flip of a single bond $V_{ij}$ can create two Ising fluxes, the entropically-favored proliferation of closed-loop domain walls at small $J$ can become strongly suppressed at large $g$ (i.e., the Monte Carlo dynamics can become effectively non-ergodic) if only the first update is employed.

Once a domain wall configuration $V$ is sampled, we employ the PyMatching python package \cite{higgott2023sparse} to compute a minimal-weight pairing of the Ising fluxes using a sparsified version of the blossom algorithm \cite{edmondsPathsTreesFlowers1965}. A priori, the algorithm developed in Ref.~\cite{higgott2023sparse} is designed for performing classical error correction: given a classical error-correcting code defined by a $\mathbb{Z}_2$-valued $k \times n$ parity-check matrix $\mathsf{H}$, the algorithm takes as input a $k$-bit ``syndrome'' $\mathsf{s} \in \text{Im}(\mathsf{H})$ and returns the smallest Hamming weight $n$-bit ``error" $\mathsf{e}$ such that $\mathsf{s} = \mathsf{H} \mathsf{e}$. In the present context, the rows (columns) of $\mathsf{H}$ are in one-to-one correspondence with the plaquettes $p$ (links $\ell$) of the square lattice, such that $H_{p\ell} = 1$ if $\ell$ is contained in the plaquette $p$; in other words, the flipped bonds $V_{ij} = -1$ and Ising fluxes $f_p = V_{ij} V_{jk} V_{k\ell} V_{\ell i} = -1$ are considered as errors and syndromes respectively in a classical error-correcting code. Given a flux configuration $f$, the algorithm returns a domain wall configuration $P$ consistent with the flux configuration, with the least number of broken bonds $P_{ij} = -1$. The configuration $P$ is then taken as our minimal-weight pairing of the fluxes.

\subsection{Free-Fermion Computation of RBIM Partition Functions}
\label{app_subsec:fermions}
In Sec.~\ref{subsec:fs_bipartiteness}, the probability of different homology classes of domain walls is mapped onto various RBIM partition functions. These can be computed exactly by several methods; the approach employed here uses Gaussian fermion techniques inspired by Ref.~\cite{bravyiEfficientAlgorithmsMaximum2014}. See also Ref.~\cite{merzTwodimensionalRandombondIsing2002} for another approach based on free fermions.

We would like to compute the following RBIM partition function:
\begin{equation}
	Z_{\RBIM} = \sum_{\sigma} \exp \qty{ \sum_{j = 1}^L \sum_{\tau = 1}^{T-1} \qty[ J^h_{\tau j} \sigma_{\tau j} \sigma_{\tau,j+1} + J^v_{\tau j} \sigma_{\tau j} \sigma_{\tau+1,j} ] }
\end{equation}
We will consider $j = 1, \ldots , L$ as a ``spatial'' coordinate and $\tau = 1, \ldots , T$ as a ``temporal'' coordinate; $J^h_{\tau j}$ and $J^v_{\tau j}$ respectively denote couplings in the horizontal (``spacelike'') and vertical (``timelike'') directions. It will prove to be convenient to work with cylindrical boundary conditions, where the spatial direction is periodic (i.e., $\sigma_{\tau,L+1} \equiv \sigma_{\tau, 1}$) and the temporal direction is open. We will also assume that the $LT$ horizontal bonds $J^h_{\tau j}$ and the $L(T-1)$ vertical bonds $J^v_{\tau j}$ have already been sampled, and our task is simply to compute $Z_{\RBIM}$ with the given couplings.

As in Ref.~\cite{schultzTwoDimensionalIsingModel1964}, we can rewrite $Z_{\RBIM}$ in terms of a product of transfer matrices as follows. Let us introduce a Hilbert space of $L$ qubits arranged in a periodic chain. We define two types of nonunitary gates $U^Z_{\tau j}$ and $U^X_{\tau j}$, given by
\begin{equation}
	U^Z_{\tau j} = e^{J^h_{\tau j} Z_j Z_{j+1}} = \cosh J^h_{\tau j} \qty[ 1 + \tanh J^h_{\tau j} Z_j Z_{j+1} ], \quad U^X_{\tau j} = e^{K^v_{\tau j}} \qty[ 1 + e^{-2K^v_{\tau j}} X_j ]
\end{equation}
where $Z_j$ and $X_j$ are the Pauli matrices acting on site $j$. Note that $U^Z_{\tau j}$ is a two-site gate, while $U^X_{\tau j}$ is a one-site gate. We further define the transfer matrices $T_Z(\tau)$ and $T_X(\tau)$:
\begin{equation}
	T_Z(\tau) = \prod_{j = 1}^L U^Z_{\tau j}, \quad T_X(\tau) = \prod_{j = 1}^L U^X_{\tau j}
\end{equation}
Finally, define $\ket{\psi_0} = 2^{L/2} \ket{+}^{\otimes L}$ to be the (unnormalized) equal superposition over all computational basis states. Putting these ingredients together, the RBIM partition function can be represented as
\begin{equation}
\label{eq:app_Z_RBIM}
	Z_{\RBIM} = \bra{\psi_0} T_Z(T) T_X(T-1) T_Z(T-1) \ldots T_X(1) T_Z(1) \ket{\psi_0}
\end{equation}
This identity is immediately verified upon inserting $T-2$ resolutions of the identity in the computational basis.

It is convenient to represent the above matrix element in terms of Majorana fermions. We define $2L$ Majorana fermions $\gamma_j$ via the following Jordan-Wigner transformation:
\begin{equation}
	\gamma_{2j-1} = \qty[\prod_{i = 1}^{j-1} X_i] Z_j, \quad \gamma_{2j} = \qty[ \prod_{i = 1}^{j-1} X_i ] Y_j
\end{equation}
In this language, $X_j = i \gamma_{2j-1} \gamma_{2j}$ and $Z_j Z_{j+1} = i \gamma_{2j} \gamma_{2j+1}$. Note that the last link in the periodic chain is given by $Z_L Z_1 = -i \Pi \gamma_{2L} \gamma_1$, where $\Pi = \prod_{i = 1}^L X_i$ is the global parity. However, since $\ket{\psi_0}$ is parity-even and all other operators commute with $\Pi$, we may freely set $\Pi = 1$. In terms of Majoranas, each of $U^Z_{\tau j}$ and $U^X_{\tau j}$ is a two-site gate, and we can think of $Z_{\RBIM}$ as a brick-wall circuit of Majorana fermions.

The important observation is that the matrix element \eqref{eq:app_Z_RBIM} can be evaluated with relative efficiency using Gaussian fermion techniques \cite{terhalClassicalSimulationNoninteractingfermion2002,bravyiLagrangianRepresentationFermionic2004}. Let us define the (unnormalized) state $\ket{\psi_{\tau}}$ via
\begin{equation}
	\ket{\psi_{\tau}} = T_Z(\tau) T_X(\tau - 1) T_Z(\tau - 1) \ldots T_X(1) T_Z(1) \ket{\psi_0}
\end{equation}
Notably, $\ket{\psi_0}$ is a Gaussian state, and each of the gates $U^Z_{\tau j}$ and $U^X_{\tau j}$ preserves Gaussianity \cite{bravyiLagrangianRepresentationFermionic2004}. Therefore, we need only to track the evolution of the correlation matrix $G_{ij}(\tau)$ and the norm $\Gamma(\tau)$ of $\ket{\psi_{\tau}}$, defined respectively by
\begin{equation}
	G_{ij}(\tau) = \frac{\bra{\psi_{\tau}} i \gamma_i \gamma_j \ket{\psi_{\tau}}}{\bra{\psi_{\tau}} \ket{\psi_{\tau}}}, \quad \Gamma(\tau) = \sqrt{\bra{\psi_{\tau}} \ket{\psi_{\tau}}}
\end{equation}
Each of these quantities can be updated gate-by-gate by using Wick's theorem. After $T$ time steps, the partition function is given by \cite{bravyiLagrangianRepresentationFermionic2004,bravyiEfficientAlgorithmsMaximum2014}
\begin{equation}
 	Z_{\RBIM} = \bra{\psi_0} \ket{\psi_T} = \Gamma(T) \det \qty[ G(T) + G(0) ]^{1/4}
\end{equation}

In principle, this algorithm is exact. However, it is empirically observed that this algorithm suffers from particularly strong numerical rounding errors within the ferromagnetic phase of the RBIM. In particular, although the covariance matrix $G$ ought to satisfy an exact orthogonality condition $G^T G = \mathds{1}$ for any pure Gaussian state, we find numerically that this condition is violated in the ferromagnetic phase of the RBIM for modestly large system sizes. As suggested by Ref.~\cite{bravyiEfficientAlgorithmsMaximum2014}, we can strongly suppress these errors by manually enforcing this orthogonality condition after each layer of gates.

\section{Additional Numerical Results in the Fradkin-Shenker Model}
\label{app:more_numerics}
In this Appendix, we provide additional numerical results for computational transitions in the Fradkin-Shenker model, for cases beyond what was considered in the main text. Specifically, we first demonstrate our minimal-weight matching protocol in the case of \textit{open} boundary conditions, where the absence of non-contractible cycles in the lattice implies that the algorithm always succeeds in constructing a well-defined order parameter. Second, we explore both minimal-weight matching and the ``intrinsic'' bipartiteness transition for additional values of $g$, showing that the qualitative behavior and quantitative values of the critical exponent remain consistent as $g$ is tuned.

\begin{figure}[t]
    \centering
    \includegraphics[width = 0.7\textwidth]{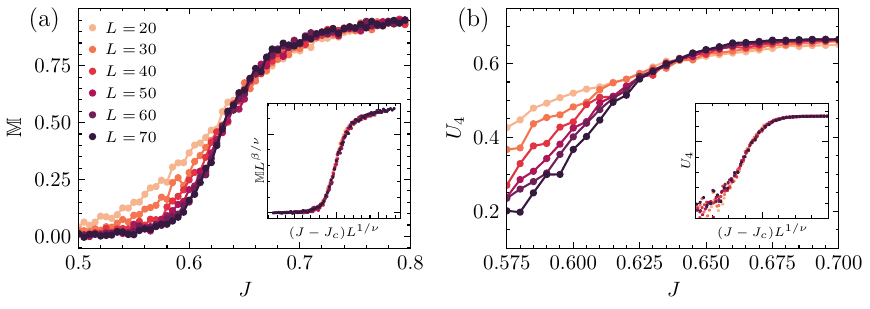}
    \caption{(a) Computational magnetization $\mathbb{M}$ in the Fradkin-Shenker model [Eq.~\eqref{eq:FS_M_comp}] established via minimal-weight pairing, for square $L \times L$ lattices with \textit{open} boundary conditions, for various linear system sizes $L$. Due to the absence of non-contractible cycles in the lattice, minimal-weight pairing always succeeds in defining an order parameter, which can then detect a computational phase transition. Inset: finite-size scaling collapse using the same estimated critical point $J_{c,\MW} \simeq 0.6298$ and estimated critical exponents $\nu_{\MW} \simeq 1.4015$ and $\beta_{\MW} \simeq 0.1096$ as in the periodic case (Fig.~\ref{fig:mwpm_numerics}). (b) Binder cumulant $U_4$ for the computational magnetization [Eq.~\ref{eq:app_binder}], which can be used to obtain estimates for $J_c$ and $\nu$ in the absence of a minimal-weight failure rate [Fig.~\ref{fig:mwpm_numerics}(c)].}
    \label{fig:openBC}
\end{figure}

\begin{figure}
    \centering
    \includegraphics[width = \textwidth]{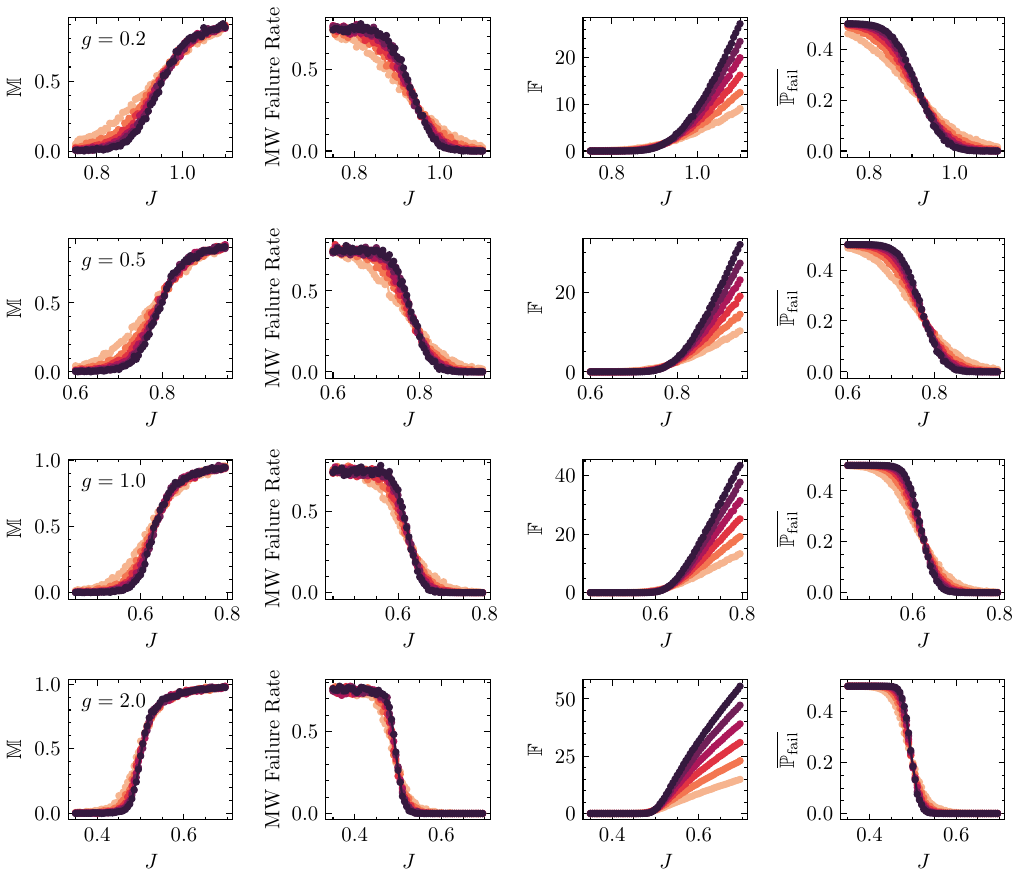}
    \caption{Computational magnetization $\mathbb{M}$, minimal-weight pairing failure rate, conditional domain wall free energy cost $\mathcal{F}$, and the ``optimal'' failure rate $P_{\text{fail}}$ in the Fradkin-Shenker model, for various values of $g$. The case $g = 1$ was considered in the main text. Each of these observables remains qualitatively similar as $g$ is tuned, with the exception of modified critical points $J_{c,\MW}$ and $J_{c,\opt}$. In the regimes shown ($g \leq 2.0$), the critical exponents $\nu_{\MW}$, $\beta_{\MW}$, $\nu_{\opt}$, and $\beta_{\opt}$ remain largely numerically stable.}
    \label{fig:other_g}
\end{figure}

We first consider the case of open boundary conditions, with numerical results shown in Fig.~\ref{fig:openBC}. In Fig.~\ref{fig:openBC}(a) we present the computational magnetization $\mathbb{M}$ as a function of $J$, with $g = 1$ fixed as in the main text. The computational magnetization exhibits an order-disorder phase transition as $J$ is tuned, demonstrating that pairing failures in the minimal-weight pairing algorithm are not necessary to achieve a computational paramagnetic phase. In the inset, we perform a finite-size scaling collapse of $\mathbb{M}$ with the critical point $J_{c,\MW}$ and exponents $\nu_{\MW}, \beta_{\MW}$ estimated from the periodic case in the main text; the excellent scaling collapse suggests that the transition in the presence of open boundary conditions is the same transition as in the case of periodic boundary conditions, despite the absence of pairing failures.

As an additional check on the consistency in critical exponents between open and periodic boundary conditions, Fig.~\ref{fig:openBC}(b) depicts the Binder parameter $U_4$ defined from moments of the computational magnetization as follows \cite{landauGuideMonteCarlo2014}:
\begin{equation}
\label{eq:app_binder}
    U_4 \equiv 1 - \frac{\expval{\mathbb{M}^4}}{3\expval{\mathbb{M}^2}^2}
\end{equation}
Since $U_4$ is expected to exhibit a scaling collapse with no scaling of the vertical axis, its computation in Monte Carlo simulations allows for the critical point $J_c$ to be determined from the crossing of different system sizes, and the correlation length exponent $\nu$ to be determined from a single-parameter scaling collapse. This was not necessary for the periodic case in the main text, where the pairing failure rate similarly required no vertical scaling and played a similar role. In the case of open boundary conditions, however, the Binder parameter is necessary to perform finite-size scaling directly on the open boundary data. This procedure results in estimated critical exponents (not shown) which are very close to those estimated from the periodic case.

In Fig.~\ref{fig:other_g}, we present the most pertinent computational observables in the Fradkin-Shenker model for both minimal-weight pairing and for the bipartiteness transition, for values of $g$ beyond the case $g = 1$ considered in the main text. We find that the behavior of these observables remains largely consistent as $g$ is tuned; in particular, they exhibit finite-size scaling collapses with numerically similar exponents for each value of $g$ considered. As can be expected, the critical points $J_{c,\MW}(g)$ and $J_{c,\opt}(g)$ shift downward as $g$ increases; in particular, both critical points approach the clean Ising value $J_c \approx 0.441$ as $g$ increases.

In the parameter ranges observed, we find (not shown) that the observed critical exponents $\nu_{\MW}$, $\beta_{\MW}$, $\nu_{\opt}$, and $\beta_{\opt}$ remain largely numerically stable. Since these exponents seem to differ from those of the clean $2d$ Ising model, it is interesting to ask how these critical exponents cross over to the clean Ising values as $g$ increases. Such questions require more detailed numerical analysis, and are beyond the scope of the present work.

\section{Thermodynamic AF/PM Hexatic Phase Transition in a Buckled Coulomb Crystal}
\label{app_buckled_crystal}
In this Appendix, we consider a slightly different model of antiferromagnetic melting than the one investigated in the main text. Namely, we consider the model proposed in Refs.~\cite{podolskyBucklingTransitionsClock2016,shamai2018molten} in which repulsively interacting ions are trapped in a two-dimensional plane via a harmonic potential in the transverse direction. When the ions are perfectly trapped in the plane, they naturally freeze into a triangular lattice; however, as the strength of the trap is reduced, the ions undergo a ``buckling transition'' in which the ions in three sublattices separate in the transverse direction \cite{podolskyBucklingTransitionsClock2016}. The $3!$ choices for how the three sublattices can separate leads to a description of the solid phase in terms of an effective six-state clock model. 

As noted in Ref.~\cite{shamai2018molten}, dislocations in the triangular lattice are energetically bound to $\pm \frac{1}{3}$ fractional vortices\footnote{Quite generally, for a $q$-state clock model on the square lattice with the Hamiltonian $H_q = - J \sum_{\expval{ij}} \cos \qty[ \frac{2\pi}{q}(n_i - n_j) ]$, a $+ \frac{k}{q}$ and $-\frac{k}{q}$ vortex are inserted at plaquettes $p$ and $p'$ by drawing a directed path $\Gamma$ from $p$ to $p'$ through the dual lattice and modifying $n_i - n_j$ to $n_i - n_j - k$ within the Hamiltonian for bonds $\expval{ij}$ which cross this path; here $i$ and $j$ are chosen so that the directed bond from $j$ to $i$ is $90^{\circ}$ counterclockwise-rotated from the directed bond in $\Gamma$ \cite{joseRenormalizationVorticesSymmetrybreaking1977}. This construction includes the familiar case of the Ising disorder operator, which can be regarded as a $\frac{1}{2}$ vortex.} of the clock model; this is a natural generalization of the binding of dislocations to Ising gauge fluxes in the model discussed throughout the main text. However, notice that in the present case, dislocations do not bind to the smallest allowed fractional vortex. As a result, we shall show that a thermodynamic phase transition between the AF and PM hexatic phases \textit{is} allowed in this case, in contrast to the case presented in the main text. 

In Sec.~\ref{sec:noPT}, we argued that an effective model for the Ising degrees of freedom within the tetratic phase was a \textit{gauged} Ising model, where each Ising gauge flux modeled the effect of a dislocation on the Ising degrees of freedom. Similarly here, it is straightforward to argue analogously that the effect of dislocations within the hexatic phase can be modeled by coupling the six-state clock model on the square lattice to a dynamical $\mathbb{Z}_3$ gauge field. Our effective lattice model is therefore given by the following Hamiltonian:
\begin{equation}
    H_{1/3}[n,U] = -J \sum_{i\mu} \cos \qty[ \frac{2\pi}{6} \qty(\Delta_{\mu} n_i - 2U_{i\mu}) ] - g \sum_{i} \cos \qty[ \frac{2\pi}{3} \varepsilon_{\mu \nu} \Delta_{\mu} U_{i\nu}   ] ,
\end{equation}
where $n_i = 0,\ldots,5$ represents the six states of the clock model, and $U_{i\mu} = 0,\pm 1$ $(\mu, \nu = x,y)$ represents a gauge field which couples only to the even part of $n_i$. For convenience, we employ common lattice gauge theory notation \cite{kogut1979introduction}: the quantity $\Delta_{\mu} n_i \equiv n_{i+\mu} - n_i$ is a lattice gradient, while $\varepsilon_{\mu \nu} \Delta_{\mu} U_{i\nu} \equiv U_{i+x,y} - U_{i,y} - U_{i+y,x} + U_{i,x}$ is a lattice curl. A $\pm \frac{1}{3}$ vortex occurs on each $i$th plaquette (i.e., the plaquette above and to the right of site $i$) whenever $\varepsilon_{\mu \nu} \Delta_{\mu} U_{i\nu} = \pm 1$ mod 3.

We shall now show that such a six-state clock model with $\pm \frac{1}{3}$ vortices is dual to a six-state clock model with an explicit symmetry-breaking field which reduces the symmetry in the dual model from $\mathbb{Z}_6$ to $\mathbb{Z}_2$. This is in contrast to the FS model investigated in the main text, where the dual model is an Ising model in a symmetry-breaking field which breaks the $\mathbb{Z}_2$ symmetry completely. As a consequence, there can exist distinct AF and PM hexatic phases in the buckled ion model which are separated by an Ising-like phase transition.

Towards this end, it is technically convenient to replace the above Hamiltonian with a Villain-type Hamiltonian containing the same symmetries \cite{joseRenormalizationVorticesSymmetrybreaking1977}. We introduce an additional integer-valued gauge field $p_{i\mu} \in \mathbb{Z}$ on the links $(i\mu)$ of the square lattice to resolve the local invariance $n_i \to n_i + 6$, and replace the cosine with a simple Gaussian Hamiltonian:
\begin{equation}
    H_{1/3, V}[n,p,U] = \frac{J_V}{2} \sum_{i\mu} \qty[ \Delta_{\mu} n_i - 2U_{i\mu} - 6 p_{i\mu} ]^2 - g \sum_{i} \cos \qty[ \frac{2\pi}{3} \varepsilon_{\mu \nu} \Delta_{\mu} U_{i\nu}   ] .
\end{equation}
Alternatively, the Poisson summation formula allows us to rewrite the partition function $Z_{1/3,V} = \sum_{n,p,U} e^{-H_{1/3,V}[n,p,U]} = \sum_{n,\ell,U} e^{-H'_{1/3,V}[n,\ell,U]}$ in terms of a Fourier-transformed representation, which replaces $p_{i\mu}$ with another integer-valued field $\ell_{i\mu}$ on the links of the lattice:
\begin{equation}
    H_{1/3,V}'[n,\ell,U] = \sum_{i\mu} \qty[ \frac{1}{2J_V} \ell_{i\mu}^2 + \frac{2\pi i}{6} \ell_{i\mu} (\Delta_{\mu} n_i - 2 U_{i\mu}) ] - g \sum_{i} \cos \qty[ \frac{2\pi}{3} \varepsilon_{\mu \nu} \Delta_{\mu} U_{i\nu}  ] .
\end{equation}
In this form the Hamiltonian is linear in the clock variables $n_i$, and they can be summed over. Performing this sum imposes a mod-6 divergenceless constraint $\Delta_{\mu} \ell_{i\mu} \equiv \ell_{ix} - \ell_{i-x} + \ell_{iy} - \ell_{i-y,y} = 0$ mod 6 on the $\ell_{i\mu}$ variables. This constraint can be resolved by writing $\ell_{i\mu} = \varepsilon_{\mu \nu} (\Delta_{\nu} m_a - 6 q_{a\nu})$, where the sites $a$ live on the dual lattice; here $m_a = 0,\ldots,5$ is a new six-state clock  variable, while $q_{a\nu} \in \mathbb{Z}$ serves as a new integer-valued Villain gauge field on the links $(a\nu)$ of the dual lattice.

If we identify the original gauge field $U_{i\mu} = \varepsilon_{\mu \nu} U_{a\nu}$ with a corresponding field $U_{a\nu}$ on the links of the dual lattice, we finally obtain
\begin{equation}
\begin{split}
    H_{1/3,V}''[m,q,U] &= \sum_{a\mu} \qty[ \frac{1}{2J_V} (\Delta_{\mu} m_a - 6 q_{a\mu})^2 - \frac{2\pi i}{3} (\Delta_{\mu} m_a) U_{a\mu} ] - g \sum_{a} \cos \qty[ \frac{2\pi}{3} \Delta_{\mu} U_{a\mu}  ] \\
    &= \sum_{a\mu} \qty[ \frac{1}{2J_V} (\Delta_{\mu} m_a - 6 q_{a\mu})^2 + \frac{2\pi i}{3} m_a \Delta_{\mu} U_{a\mu} ] - g \sum_{a} \cos \qty[ \frac{2\pi}{3} \Delta_{\mu} U_{a\mu}  ] .
\end{split}    
\end{equation}
Notice that the quantity $\Delta_{\mu} U_{a\mu} = \varepsilon_{\mu \nu} \Delta_{\mu} U_{i\nu}$ is exactly the gauge flux through the original lattice's plaquettes. Whenever a $\pm \frac{1}{3}$ vortex is present at the  dual lattice site $a$, a factor of $\exp{ \mp \frac{2\pi i}{3} m_a - \frac{3}{2}g }$ is inserted into the partition function. By treating the vorticity at each dual lattice site independently\footnote{Note that in a periodic system, the global vorticity must sum to an integer; for example, one can have one $+\frac{1}{3}$ vortex and one $-\frac{1}{3}$ vortex, or three $+\frac{1}{3}$ vortices, but not just one $\frac{1}{3}$ vortex. This constraint is nicely handled by the corresponding insertions of the factors $e^{2 \pi i m_a / 3}$, resulting in a a correlation function which vanishes by symmetry unless the total number of these insertions adds to a multiple of 3. Thus, for the purposes of integrating out $U_{a\mu}$ in this step, there is no global issue with treating the vorticity in each plaquette as independent.}, we can finally integrate out $U_{a\mu}$ to obtain the final Hamiltonian
\begin{equation}
    H'''_{1/3,V}[m,q] = \frac{1}{2J_V} \sum_{a\mu} \qty(\Delta_{\mu} m_a - 6 q_{a\mu})^2 - h(g) \sum_a \cos \qty( \frac{2\pi}{3} m_a) , \quad h(g) = \frac{2}{3} \log \qty{ \frac{1 + 2e^{-3g/2}}{1 - e^{-3g/2}}} ,
\end{equation}
where $h(g)$ is a positive and monotonically decreasing function of $g$. 

Altogether, we find that a six-state Villain clock model with $\pm \frac{1}{3}$ vortices is exactly dual to a six-state Villain clock model in the presence of a field which explicitly breaks the $\mathbb{Z}_6$ symmetry down to a residual $\mathbb{Z}_2$ symmetry. Thus, the presence of $\pm \frac{1}{3}$ vortices eliminates the gapless phase of the clock model, but still allows for a symmetry-breaking phase transition in the Ising universality class. This suggests that a sharp thermodynamic phase transition can occur between the AF and PM hexatic phases.

\newpage
\twocolumngrid

\bibliographystyle{apsrev4-2-author-truncate}
\bibliography{refs}

\end{document}